\documentclass[12pt]{iopart}
\usepackage{graphicx}
\usepackage{iopams}

\begin{document}

\title{Astrophysical and cosmological probes of dark matter}

\author{Matts Roos}
\address{Department of Physics, FI-00014 University of Helsinki, Finland}
\ead{matts.roos@helsinki.fi}
\date{\today}

\begin{abstract}
Dark matter has been introduced to explain substantial mass deficits noted at different astronomical scales, in galaxies, groups of galaxies, clusters, superclusters and even across the full horizon. Dark matter does not interact with baryonic matter except gravitationally, and therefore its effects are sensed only on the largest scales. Although it is still unknown whether dark matter consists of particles or of a field or has some other nature, it has a rich phenomenology. This review summarizes all the astrophysical and cosmological probes that have produced overwhelming evidence for its existence. The breadth of the subject does not permit details on the observational methods (the reference list then helps), thus the review is intended to be useful mainly to cosmologists searching to model dark matter.

\end{abstract}
 \maketitle

\section*{Contents}
1. Introduction\\
2. Stars near the Galactic disk\\
3. Virially bound systems\\
4. Rotation curves of spiral galaxies\\
5. Strong and weak lensing\\
6. Elliptical galaxies\\
7. Mass-to-light ratios and dwarf spheroidals\\
8. Small galaxy groups emitting X-rays\\
9. Mass autocorrelation functions\\
10. Cosmic Microwave Background\\
11. Baryonic Acoustic Oscillations\\
12. Galaxy formation in purely baryonic matter?\\
13. Large Scale Structures simulated\\
14. Dark matter from overall fits\\
15. Merging galaxy clusters\\
16. Comments and conclusions

\section{Introduction}

Apparently the matter content of the Universe is dominated by an unknown form of dark matter (DM) without interactions with ordinary baryonic matter, perhaps not even with itself. It only interacts via the gravitational field, manifesting its effects on astrophysical and cosmological scales. The purpose of this review is to summarize the phenomenology of all such effects, that can serve as probes of dark matter. Regardless of the ultimate, correct explanation of its particle nature or field nature, theory needs to address all these effects.

This review does not cover the historical development, except by glimpses, because the rapid development of observational means tends to render all discoveries older than a decade unimportant.

Beginning from the first controversial conclusions from the motion of stars near the Galactic disk on missing matter in the Galactic disk (Sec. 2), and that of Fritz Zwicky in 1933 \cite{Zwicky} of missing matter in the Coma cluster (Sec. 3), we describe the kinematics of virially bound systems (Sec. 3) and rotating spiral galaxies (Sec. 4). An increasingly important method to determine the weights of galaxies, clusters and gravitational fields at large, independently of electromagnetic radiation, is lensing, strong as well as weak (Sec. 5). Next follows a discussion of dark matter in elliptical galaxies (Sec. 6) and mass-to-light ratios which probe dark matter in all systems, notably in dwarf spheroidals (Sec. 7). Different ways to measure missing mass in groups and clusters derive from the comparison of visible light and X-rays (Sec. 8). Mass autocorrelation functions relate galaxy masses to dark halo masses (Sec. 9).

In radiation the most important tools are the temperature and polarization anisotropies in the Cosmic Microwave Background (CMB) (Sec. 10), which give information on the mean density of both dark and baryonic matter as well as on the geometry of the Universe. The large scale structures of matter exhibit similar fluctuations evident in the Baryonic Acoustic Oscillations (BAO) (Sec. 11). The amplitude of the temperature variations in the CMB prove, that galaxies could not have formed in a purely baryonic Universe (Sec. 12). Simulations of large scale structures also show that DM must be present (Sec. 13). The best quantitative estimates of the density of DM come from overall parametric fits to cosmological models, notably the Cold Dark Matter model '$\Lambda$CDM' with a cosmological constant $\Lambda$, of CMB data, BAO data, and redshifts of supernovae of type Ia (SNe\,Ia) (Sec. 14). A particularly impressive testimony comes from merging clusters (Sec. 15). We conclude this review with a brief summary (Sec. 16).

\section{Stars near the Galactic disk}
In 1922 the Dutch astronomer Jacobus Kapteyn \cite{Kapteyn} studied the vertical motions
of all known stars near the Galactic plane and used these data to calculate the acceleration of matter. This amounts to treating the stars as members of a "star atmosphere", a statistical ensemble in which the density of stars and their velocity dispersion defines a "temperature" from which one obtains the gravitational potential.  This is analogous to how one obtains the gravitational potential of the Earth from a study of the atmosphere. Kapteyn found that the spatial density is sufficient to explain the vertical motions.

Later in the same year the British astronomer James Jeans \cite{Jeans} reanalyzed Kapteyn's data and found a mass deficit: to each bright star two dark stars had to be present.
The result contradicted grossly the expectations: if the potential provided by the known stars was not sufficient to keep the stars bound to the Galactic disk, the Galaxy should rapidly be losing stars. Since the Galaxy appeared to be stable there had to be some missing matter near the Galactic plane.

In 1932 the Dutch astronomer Jan Hendrik Oort \cite{Oort1} reanalyzed the vertical motions and came to the same conclusion as Jeans. There was indeed a mass deficit which Oort proposed to indicate the presence of some dark matter in our Galaxy. The possibility that this missing matter would be nonbaryonic could not even be thought of at that time. Note that the first neutral baryon, the neutron, was discovered by James Chadwick \cite{Chadwick} only in the same year, in 1932.

However, it is nowadays considered, that this does not prove the existence of DM in the disk. The potential in which the stars are moving is not only due to the disk, but rather to the totality of matter in the Galaxy which is dominated by the Galactic halo. The advent of much more precise data in 1998 led Holmberg \& Flynn  \cite{Holmberg} to conclude that no DM was present in the disk.

Oort determined the mass of the Galaxy to be $10^{11}~\rm M_{sun}$, and thought that the nonluminous component was mainly gas. Still in 1969 he thought that intergalactic gas  made up a large fraction of the mass of the universe \cite{Oort2}. The general recognition of the missing matter as a possibly new type of non-baryonic DM dates to the early eighties.

\section{Virially bound systems}
The planets move around the Sun along their orbits with orbital velocities balanced by the total gravity of the Solar system. Similarly, stars move in galaxies in orbits with orbital velocities $v$ determined by the gravitational field of the galaxy, or they move with velocity dispersion $\sigma$. Galaxies in turn move with velocity dispersion $\sigma$ under the influence of the gravitational field of their environment, which may be a galaxy group, a cluster or a supercluster. In the simplest dynamical framework one treats massive systems (galaxies, groups and clusters) as statistically steady, spherical, self-gravitating systems of $N$ objects with average mass $m$ and average velocity $v$ or velocity dispersion $\sigma$. The total kinetic energy $E$ of such a system is then (we now use $\sigma$ rather than $v$)

\begin{equation}
E=(1/2)Nm\sigma^2\ .
\end{equation}
If the average separation is $r$, the potential energy of  $N(N-1)/2$ pairings is
\begin{equation}
U=-(1/2)N(N-1)Gm^2/r\ .
\end{equation}
The \textit{virial theorem} states that for such a system
\begin{equation}
E=-U/2\ .\label{vir}
\end{equation}
The total dynamic mass $M_{dyn}$ can then be estimated from $\sigma$ and $r$
\begin{equation}
M_{dyn}=Nm=2r\sigma^2/G\ .
\end{equation}
This can also be written
\begin{equation}
\sigma^2\propto(M_{dyn}/L)IR\ ,
\end{equation}
where $I$ is a surface luminosity, {R} is a scale, and $M_{dyn}/L$ is the \textit{mass-to-light ratio}. Choosing the scale to be the half light radius $R_e$, this implies a relationship between the observed central velocity dispersion $\sigma_0$, $I_e$ and $R_e$ called the \textit{Fundamental Plane}. of the form
\begin{equation}
R_e\propto(\sigma_0)^a (I_e)^b\ .\label{FundPlane}
\end{equation}
The virial theorem predicts the values $a=2$, $b=1$ for the coefficients. This relationship is found in ellipticals \cite{Jorgensen, Hyde} and in some other types of stellar populations, but with somewhat different coefficients.

\subsection{Halo density profiles}

The shapes of DM halos in galaxies and clusters need to be simulated or fitted by empirical formulae. Mostly the shape is taken to be spherically symmetric so that
the total gravitating mass profile $M(r)$ depends on three parameters: the mass proportion in stars, the halo mass and the length scale. A frequently used radial density profile parametrization is
\begin{equation}
\rho_{DM}(r)=\rho_0/[(r/r_s)^{\alpha}(1+r/r_s)^{3-\alpha}]\ ,\label{halo}
\end{equation}
where $\rho_0$ is a normalization constant and $0\leq\alpha\leq 3/2$. Standard choices are $\alpha=1$ for the Navarro-Frenk-White profile (NFW) \cite{NFW}, and $\alpha=3/2$ for the profile of Moore \& al. \cite{Moore}, both \textit{cusped} at $r=0$.

Another parametrization is the Einasto profile (\cite{Einasto} and earlier references therein)
\begin{equation}
\rho_{DM}(r)=\rho_e \exp\{-d_n[(r/r_e)^{1/n}-1]\},\label{Einasto}
\end{equation}
where the term $d_n$ is a function of $n$ such that $\rho_e$ is the density at $r_e$, which defines a volume containing half of the total mass. At $r=0$ the density is then finite and \textit{cored}.

The Burkert profile \cite{Burkert} has a constant density core
\begin{equation}
\rho_{DM}(r)=\rho_0/[(1+r/r_s)(1+(r/r_s)^2)]\ ,\label{Burkert}
\end{equation}
which fitted dwarf galaxy halos well in 1995, but no longer does so, see Sec. 7.

Some clusters are not well fitted by any spherical approximation. The halo may exhibit a strong ellipticity or triaxiality in which case none of the above profiles is good.

The dependence of the physical size of clusters on the mass, characterized by the mass concentration index $c\equiv r_{vir}/r_s$, has been studied in $\Lambda$CDM simulations \cite{Klypin}. At intermediate radii $c$ is a crucial quantity in determining the density shape.

   \begin{figure}[htbp]
\includegraphics[width=12cm]{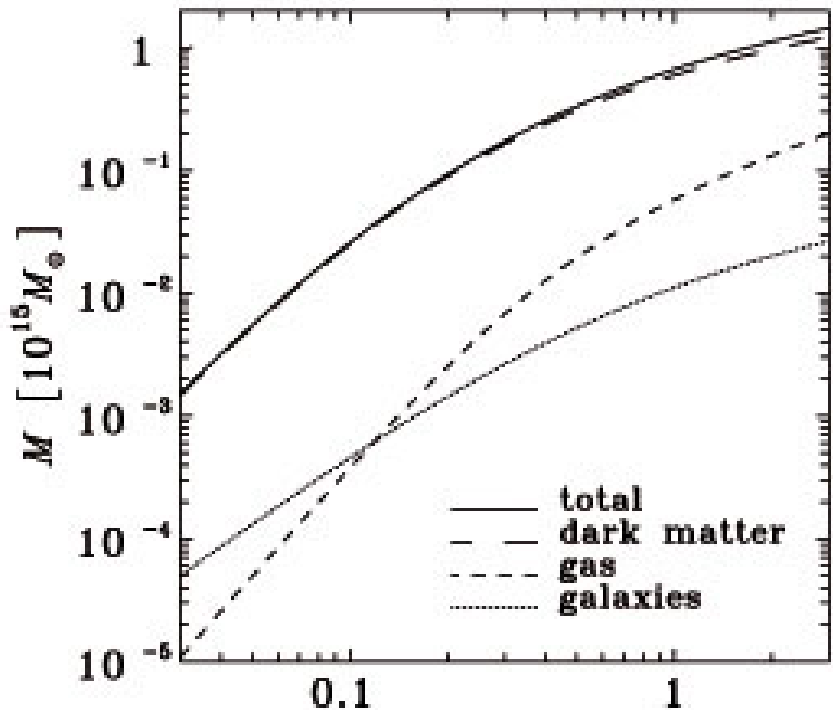}
\caption{Density profile of matter components enclosed within a given radius $r$ in the Coma cluster, versus $r/r_{vir}$. From E. L. Lokas \&  G. A. Mamon \cite{Lokas}.}
\end{figure}

\subsection{The Coma cluster}
Historically, the first observation of dark matter in an object at a cosmological distance was made by Fritz Zwicky in 1933 \cite{Zwicky}. While measuring radial velocity dispersions of member galaxies in the Coma cluster (that contains some $1000$ galaxies), and the cluster radius from the volume they occupy, Zwicky was the first to use the virial theorem to infer the existence of unseen matter. He found to his surprise that the dispersions were almost a factor of ten larger than expected from the summed mass of all visually observed galaxies in
the Coma. He concluded that in order to hold galaxies together the cluster must contain huge amounts of some non-luminous matter. From the dispersions he concluded that the average mass of
galaxies within the cluster was about 160 times greater than expected from their luminosity (a value revised today), and he proposed that most of the missing matter was dark.

Zwicky's suggestion was not taken seriously at first by the astronomical community which Zwicky felt as hostile and prejudicial. Clearly, there was no candidate for the dark matter because gas radiating X-rays and dust radiating in the infrared could not yet be observed, and non-baryonic matter was unthinkable.  Only some forty years later when studies of motions of stars within galaxies also implied the presence of a large halo of unseen matter extending beyond the visible stars, dark matter became a serious possibility.

Since that time, modern observations have revised our understanding of the composition of clusters. Luminous stars represent a very small fraction of a cluster mass; in addition there is a baryonic, hot \textit{intracluster medium} (ICM) visible in the X-ray spectrum. Rich clusters typically have more mass in hot gas than in stars; in the largest virial systems like the Coma the composition is about 85\% DM, 14\% ICM, and only 1\% stars \cite{Lokas}.

In modern applications of the virial theorem one also needs to model and parametrize the radial distributions of the ICM and the dark matter densities. In the outskirts of galaxy clusters the virial radius roughly separates bound galaxies from galaxies which may either be infalling or unbound. The virial radius $r_{vir}$ is conventionally defined as the radius within which the mean density is 200 times the background density.

Matter accretion is in general quite well described within the approximation of the \textit{Spherical Collapse Model}. According to this model, the velocity of the infall motion and the matter overdensity are related. Mass profile estimation is thus possible once the infall pattern of galaxies is known \cite{Cupani}.

In Fig. 1 the Coma profile is fitted \cite{Lokas} with Eq.~(\ref{halo}) with $\alpha=0$ which describes a centrally finite profile which is almost flat. The separation of different components in the core is not well done with Eq.~(\ref{halo}) because the Coma has a binary center like many other clusters \cite{Coma}.

   \begin{figure}[htbp]
\includegraphics[width=12cm]{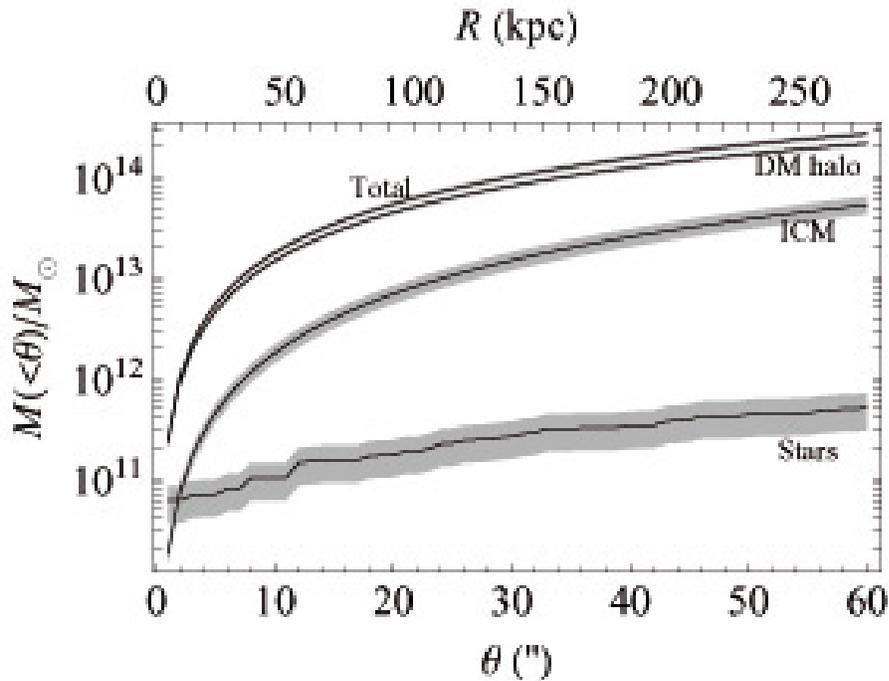}
\caption{Density profile of matter components in the cluster AC 114, enclosed within a given projected radius. From M. Sereno \& {\em al.} \cite{Sereno}.}
\end{figure}

\subsection{The AC 114 cluster}
Dark matter is usually dissected from baryons in lensing analyses by first fitting the lensing features to obtain a map of the total matter distribution and then subtracting the gas mass fraction as inferred from X-ray observations \cite{Bradac, Allen}.
The total mass map can then be obtained with parametric models in which the contribution from cluster-sized DM halos is considered together with the main galactic DM halos \cite{Limousin}. Mass in stars and in stellar remnants is  estimated converting galaxy luminosity to mass assuming suitable stellar mass to light ratios.

One may go one step further by exploiting a parametric model which has three kinds of components: cluster-sized DM halos, galaxy-sized (dark plus stellar) matter halos, and a cluster-sized gas distribution \cite{Coma, Sereno}. As an example we show the results of such an analysis of the dynamically active cluster AC 114 in Fig. 2.

In systems of merging clusters DM may become spatially segregated from baryonic matter and thus observable. We shall meet several such cases in Sec. 15.

\subsection{The Local Group}
The Local Group is a very small virial system, dominated by two large galaxies, the M31 or Andromeda galaxy, and the Milky Way. The M31 exhibits blueshift, falling in towards us. Evidently our Galaxy and M31 form a bound system together with all or most of the minor galaxies in the Local Group. The Local Group extends to about 3 Mpc and the velocity dispersions of its members is about 200 km s$^{-1}$.

In this group the two large galaxies dominate the dynamics, so that it is not meaningful to define a statistically average pairwise separation between galaxies, nor an  average mass nor an average orbital velocity. The total kinetic energy $E$ is still given by the sum of all the group members, and the potential energy $U$ by the sum of all the galaxy pairs, but here the pair formed by the M31 and the Milky Way dominates, and the pairings of the smaller members with each other are negligible.

An interesting recent claim is, that the mass estimate of the Local Group is also affected by the accelerated expansion, the ``dark energy''. A. D. Chernin \& al. \cite{Valtonen} have shown that the potential energy $U$ is reduced in the force field of dark energy, so that the virial theorem for $N$ masses $m_i$ with baryocentric radius vectors ${\bf r}_i$ takes the form
\begin{equation}
E=-(1/2)U + U_2\ ,
\end{equation}
where $U$ is defined as in Eq.~(\ref{vir}), and
\begin{equation}
U_2=-(4\pi\rho_v/3)~\Sigma m_i r_i^2
\end{equation}
is a correction which reduces the potential energy due to the
background dark energy density $\rho_v$. In the Local Group this correction to the mass appears to be quite substantial, of the order of $30\% -50\%$.

The dynamical mass of the local group is $3.2 - 3.7\times 10^{12}$ solar masses whereas the total visible mass of the Galaxy + M31 is only $2\times 10^{11}$
solar masses. Thus there is a large amount of dark matter missing.

\subsection{The local Universe}
In a large volume beyond the local group, Tully in 1984 \cite{Tully} measured the velocities of 2367 galaxies with radial velocities below 3000 km s$^{-1}$. He found that the mass density parameter (which is normalized to the critical mass) in this ``Local Universe" was $\Omega_m=0.08$, in clear conflict with the global value, $\Omega_{m,global}= 0.27 \pm 0.02$ (as we shall see in Sec. 14).

More recently Karachentsev \cite{Karachentsev} has extended this analysis out to a volume of a diameter of 96 Mpc, containing 11\,000 galaxies appearing single, in pairs, in triplets and in groups. Most of them belong to the Local Supercluster and constitute $<15\%$ of the mass of Virgo. The radial velocities are $v<3500$ km s$^{-1}$. These galaxies can be treated as a virial system with average density $\Omega_{m,local}= 0.08 \pm 0.02$, again surprisingly small compared to the global density. Karachentsev quotes three proposed explanations for this mass deficit.\\
\indent -- Dark matter in the systems of galaxies extends far beyond their virial radius, so that the total mass of a group or cluster is 3 -- 4 times larger than the virial estimate. However, this contradicts other existing data.\\
\indent -- The diameter of the considered region of the Local universe, 90 Mpc, does not correspond to the true scale of the ``homogeneity cell"; our Galaxy may be located inside a giant void sized about 100 -- 500 Mpc, where the mean density of matter is 3 to 4 times lower than the global value. However, the location of our Galaxy is characterized by an excess, rather than by a deficiency of local density at all scales up to 45 Mpc.\\
\indent -- Most of the dark matter in the Universe, or about two thirds of it, is not associated with groups and clusters of galaxies, but distributed in the space between them in the form of massive dark clumps or as a smooth ``ocean". It is as yet difficult to evaluate this proposal.

Clearly the physics in the Local Universe does not prove the existence of dark matter, rather it brings in new problems.

\section{Rotation curves of spiral galaxies}

Spiral galaxies are stable gravitationally bound systems in which visible matter is composed of stars and interstellar gas. Most of the observable matter is in a relatively thin disc, where stars and gas rotate around the galactic center on
nearly circular orbits. The galaxy kinematics is measured by the Doppler shift of well-known emission lines of particular tracers of the gravitational potential: HI, CO and H$_{\alpha}$.

If the circular velocity at
radius $r$ is $v$ in a rotating galaxy with mass~$M(r)$ inside $r$, the condition for stability is that the centrifugal acceleration $v/r$ should equal the gravitational pull $GM(r)/r^2$, and the radial dependence of $v$ would then be expected to follow Kepler's law
\begin{equation}
v^2=GM(r)/r.\label{disk}
\end{equation}

The surprising result for spiral galaxy rotation curves is, that the velocity does not follow Kepler's inverse-root law, but stays rather constant after attaining a maximum. The most obvious solution to this is that the galaxies
are embedded in extensive, diffuse halos of dark matter  If the
mass $M(r)$ enclosed inside the radius $r$, is proportional to $r$
it follows that $v(r)\approx$ constant.
   \begin{figure}[htbp]
\includegraphics[width=15cm]{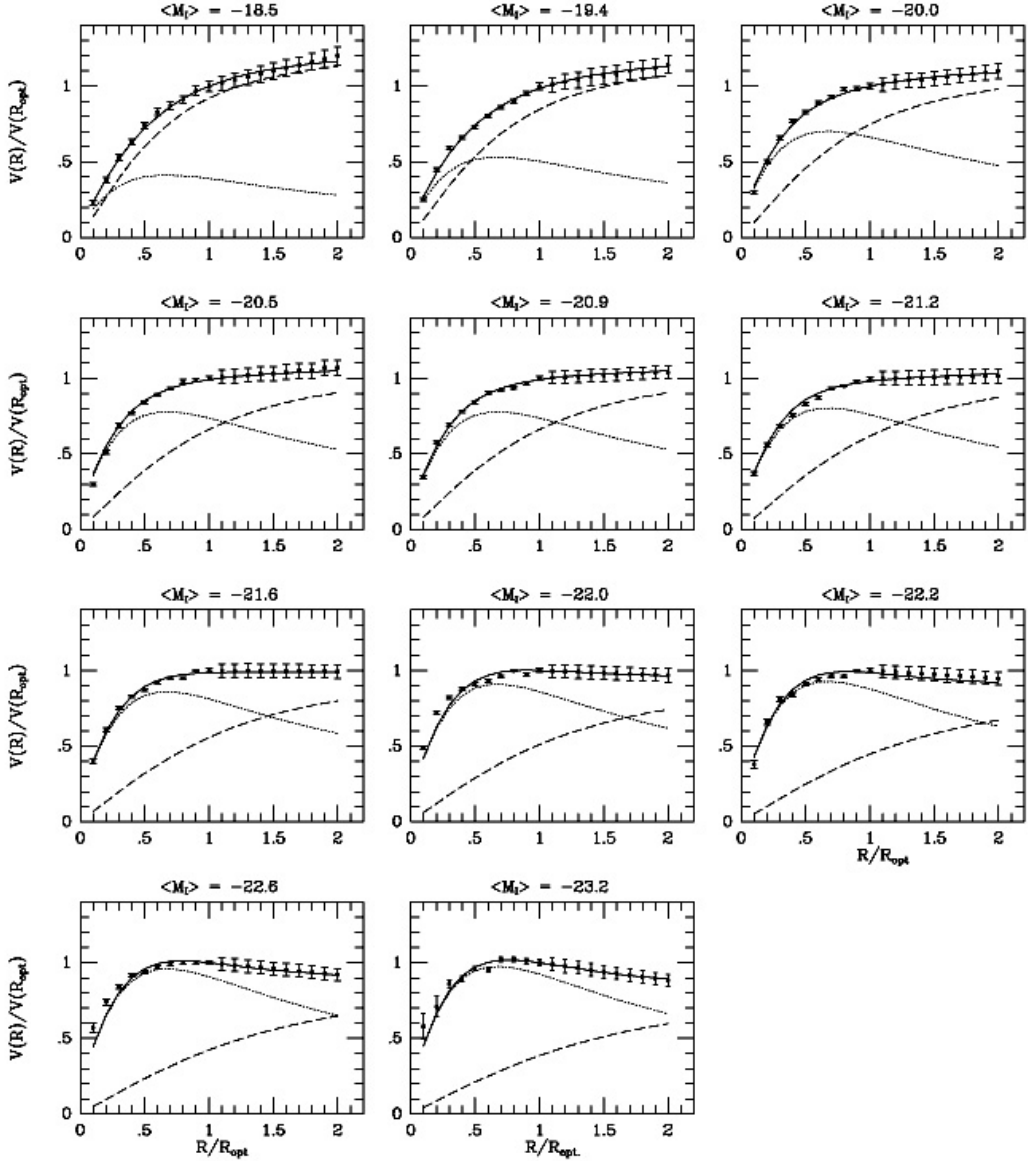}
\caption{Best disk -- halo fits to the Universal Rotation Curve (dotted line is disk, dashed line is halo). Each object is identified by the halo virial mass,increasing downwards. From P. Salucci \& $al.$ \cite{Salucci}.}
\end{figure}

The rotation curve of most galaxies can be fitted by the superposition of contributions from the stellar and gaseous disks, sometimes
a bulge, and the dark halo, modeled by a quasi -- isothermal sphere. The inner part is difficult to model because the density of stars is high, rendering observations of individual star velocities difficult. Thus the fits are not unique, the relative contributions of disk and dark matter halo is model-dependent, and it is sometimes not even sure whether galactic disks do contain dark matter. Typically, dark matter constitutes about half of the total mass.

\begin{figure}[htbp]
   \includegraphics[width=15cm]{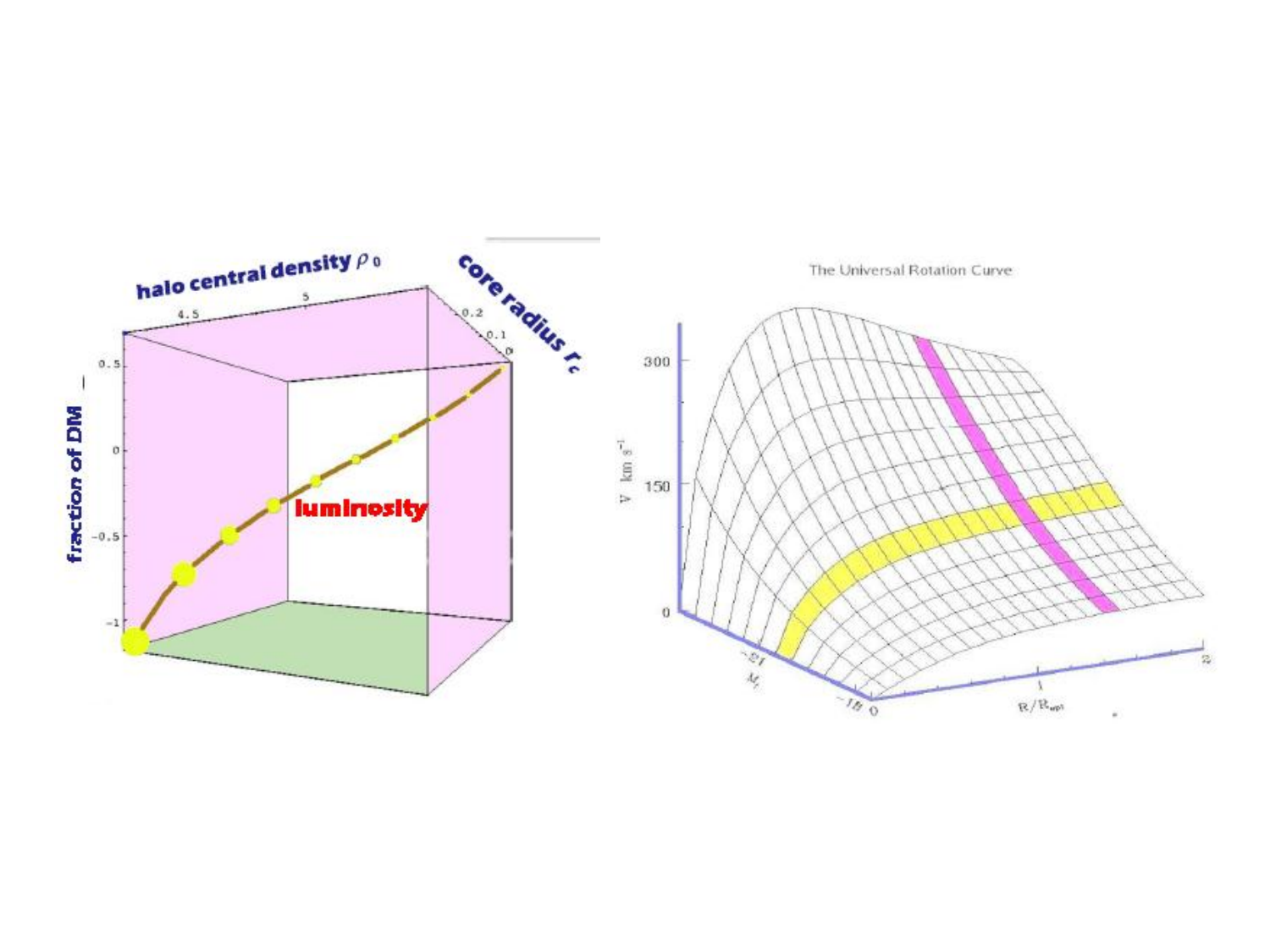}
   \caption{Left: The 4-dimensional space of luminosity, core radius, halo central density and fraction of DM. Right: The smooth surface of spiral galaxy rotation curves in the space of normalized radius $R/R_{optical}$, magnitude $M$ and rotation velocity $V$ in km s$^{-1}$. P. Salucci priv. comm. and ref. \cite{Salucci}.}
      \end{figure}

In Fig. 3 we show the rotation curves fitted for eleven well-measured galaxies \cite{Salucci}  of increasing halo mass. One notes, that the central dark halo component is indeed much smaller than the luminous disk component. At large radii, however, the need for a DM halo is obvious. On galactic scales, the contribution of DM generally dominates the total mass. Note the contribution of the baryonic component, negligible for light masses but increasingly important in the larger structures.

The mass discrepancy emerges also as a disagreement between light and mass distributions: light does not trace mass, the ratio
\begin{equation}
(dM/dr)/(dL/dr)
\end{equation}
is not constant, but increases with radius \cite{Bosma}.

Gentile \& al. \cite{Gentile} have shown that cusped profiles are in clear conflict with data on spiral galaxies. Central densities are rather flat, scaling approximately as $\rho_0\propto r^{-2/3}_{luminous}$. The best-fit disk + NFW halo mass model fits the rotation curves poorly, it implies an implausibly low stellar mass-to-light ratio and an unphysically high halo mass. Clearly the actual profiles are of very uncertain origin.

One notes in Fig.3 that the shape of the rotation curve depends on the halo virial mass so that the distribution of gravitating matter, unlike luminous matter,
is luminosity dependent. The old idea that the rotation curve stays constant after attaining a maximum is thus a simplification of the real situation.
The rotation velocity can be expressed by a \textit{Universal Rotation Curve} \cite{Salucci}. All spiral galaxies lie on a curve in the 4-dimensional space of luminosity, core radius, halo central density and fraction of DM, see Fig. 4

\begin{figure}[htbp]
   \includegraphics[width=15cm]{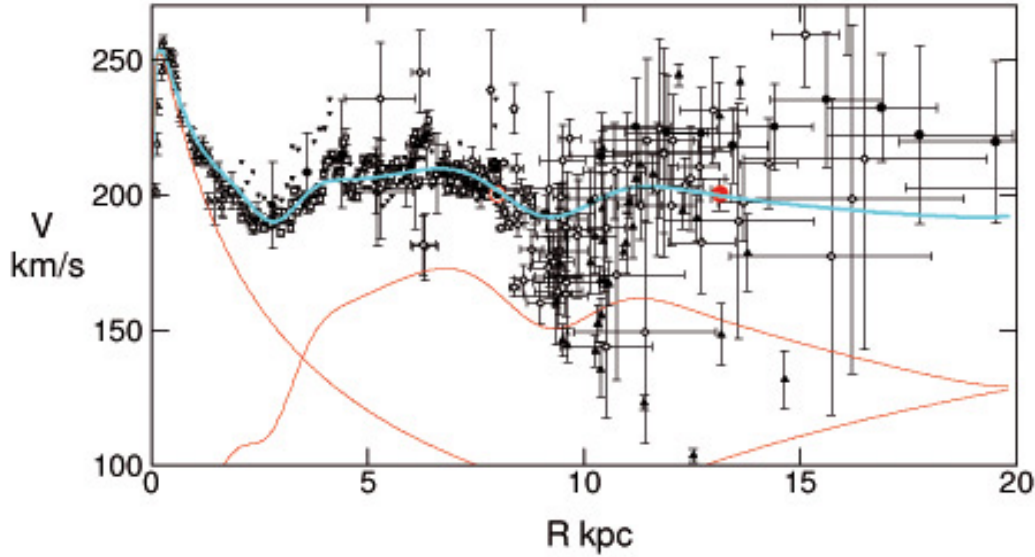}
   \caption{Decomposition of the rotation curve of the Milky Way into the components bulge, stellar disk + interstellar gas, DM halo (the red curves from left to right). From Y. Sofue {\em et al.} \cite{Sofue}.}
      \end{figure}

Our Galaxy is complicated because of what appears to be a noticeable density dip at 9 kpc and a smaller dip at 3 kpc, as is seen in Fig. 5  \cite{Sofue}. To fit the measured rotation curve one needs at least three contributing components: a central bulge, the star disk + gas, and a DM  halo \cite{Sofue, Ullio, Weber}.  For small radii there is a choice of empirical rotation curves, and no DM component appears to be needed until radii beyond 15 kpc.

\section{Strong and weak lensing}

A consequence of the \textit{Strong Equivalence Principle} (SEP) is that a photon in a gravitational field moves as if it possessed mass, and light rays therefore bend around gravitating masses. Thus celestial bodies can serve as \textit{gravitational lenses} probing the gravitational field, whether baryonic or dark without distinction.

Since photons are neither emitted nor absorbed in the process of gravitational light deflection, the surface brightness of lensed sources remains unchanged. Changing the size of the cross-section of a light bundle only changes the flux observed from a source and magnifies it at fixed surface-brightness level. If the mass of the lensing object is very small, one will merely observe a magnification of the brightness of the
lensed object an effect called \textit{microlensing}. Microlensing of distant quasars by compact lensing objects (stars, planets) has also been observed and used for estimating the mass distribution of the lens--quasar systems.

\begin{figure}[htbp]
   \includegraphics[width=12cm]{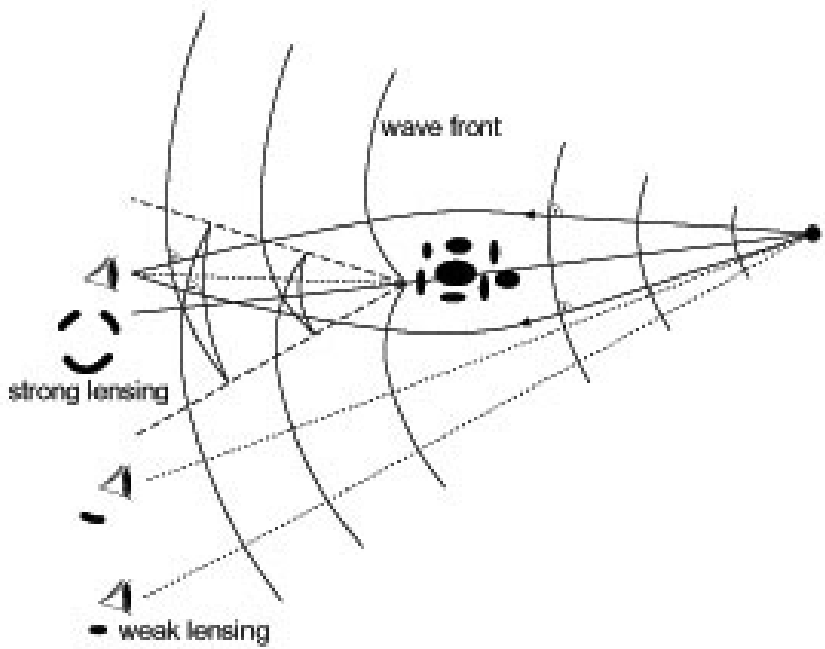}
   \caption{Wave fronts and light rays in the presence of a cluster perturbation. From N. Straumann \cite{Straumann}.}
      \end{figure}

In \textit{Strong
Lensing} the photons move along geodesics in a strong gravitational potential which distorts space as well as time, causing larger deflection angles and requiring the full theory of General Relativity. The images in the observer plane can then become quite complicated because there may be more than one null geodesic connecting source and observer; it may not even be possible to find a unique mapping onto the source plane {\em cf} Fig.6 . Strong lensing is a tool for testing the distribution of mass in the lens rather than purely a tool for testing General Relativity. An illustration is seen in Fig. 7 where the lens is an elliptical galaxy \cite{Smith}.

At cosmological distances one may observe lensing by composed objects such as galaxy groups which are ensembles of ``point-like'', individual galaxies. Lensing effects are very model-dependent, so to learn the true magnification effect one
needs very detailed information on the structure of the lens.

\begin{figure}[htbp]
\includegraphics[width=9cm]{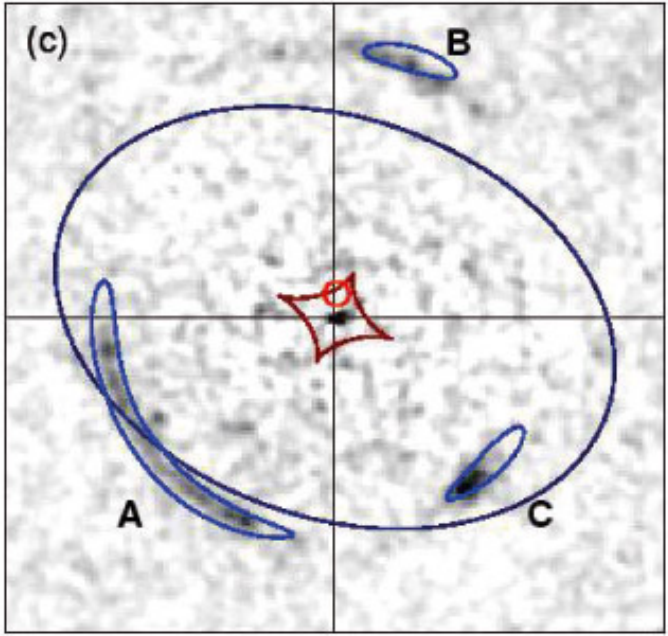}
   \caption{This image resulted from color-subtraction of a lensing singular isothermal elliptical galaxy. The strongly lensed object forms two prominent arcs A, B and a less extended third image C. From R.J. Smith \& {\em al.} \cite{Smith}}
 \end{figure}

\textit{Weak Lensing} refers to deflection through a small angle when the light ray can be treated as a straight line (Fig. 6), and the deflection as if it occurred discontinuously at the point of closest approach (the thin-lens approximation in optics). One then only invokes SEP to account for the distortion of clock rates.

 \begin{figure}[htbp]
   \includegraphics[width=11cm]{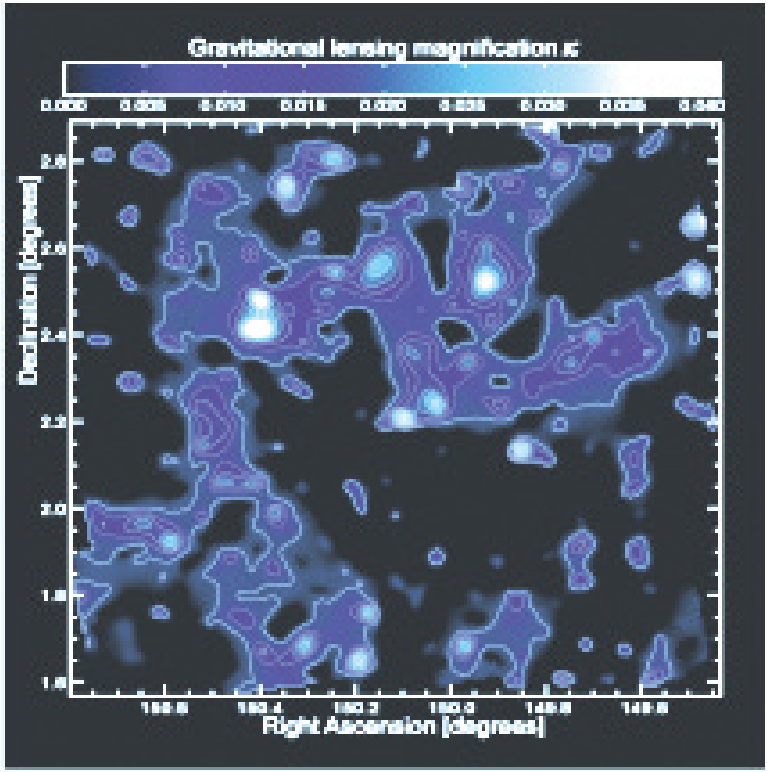}
   \caption{Map of the dark matter distribution in the 2-square degree
COSMOS field: the linear blue scale on top shows the gravitational lensing magnification $\kappa$, which is proportional to the projected mass along the line of sight. From R. Massey \& $al.$ \cite{Massey}}
\end{figure}

The large-scale distribution of matter in the Universe is inhomogeneous in every direction, so one can expect that everything we observe is displaced and distorted by weak lensing. Since the tidal gravitational field and the
deflection angles depend neither on the nature of the matter nor on its physical state, light deflection probes the total projected mass distribution. Lensing in infrared light offers an additional advantage of being able to sense distant background galaxies, since their number density is higher than in the optical range.

Background galaxies would be ideal tracers of distortions if they were intrinsically circular, because lensing transforms circular sources into ellipses. Any measured ellipticity would then directly reflect the action of the gravitational tidal field of the interposed lensing matter, and the statistical properties of the distortions would reflect the properties of the matter distribution. But many galaxies are actually intrinsically elliptical, and the ellipses are randomly oriented. This introduces noise into the inference of the tidal field from observed ellipticities. A useful feature in the sky is a fine-grained pattern of faint and distant blue galaxies appearing as a `wall paper'. This makes statistical weak-lensing studies possible, because it allows the detection of the coherent distortions imprinted by gravitational lensing on the images of the galaxy population.

Thus weak lensing has become an important technique to map non-luminous matter. A reconstruction of one of the largest and most detailed weak lensing surveys undertaken with the Hubble Space Telescope is shown in Fig. 8 \cite{Massey}. This map covers a large enough area to see extended filamentary structures.

A very large review on lensing by R. Massey \& al. \cite{Massey1} can be recommended. We show several examples of lensing by clusters in Sec. 15

\section{Elliptical galaxies}

Elliptical galaxies are quite compact objects which mostly do not rotate so their mass cannot be derived from rotation curves. The total dynamical mass is then the virial mass as derived from the velocity dispersions of stars and the anisotropies of their orbits. However, to disentangle the total mass profile into its dark and its stellar components is not straightforward, because the dynamical mass decomposition of dispersions is not unique. The luminous matter in the form of visible stars is a crucial quantity, indispensable to infer the dark component. When available one also makes use of strong and weak lensing data, and of the X-ray properties of the emitting hot gas. The gravity is then balanced by pressure gradients as given by Jeans' Equation.

 \begin{figure}[htbp]
   \includegraphics[width=15cm]{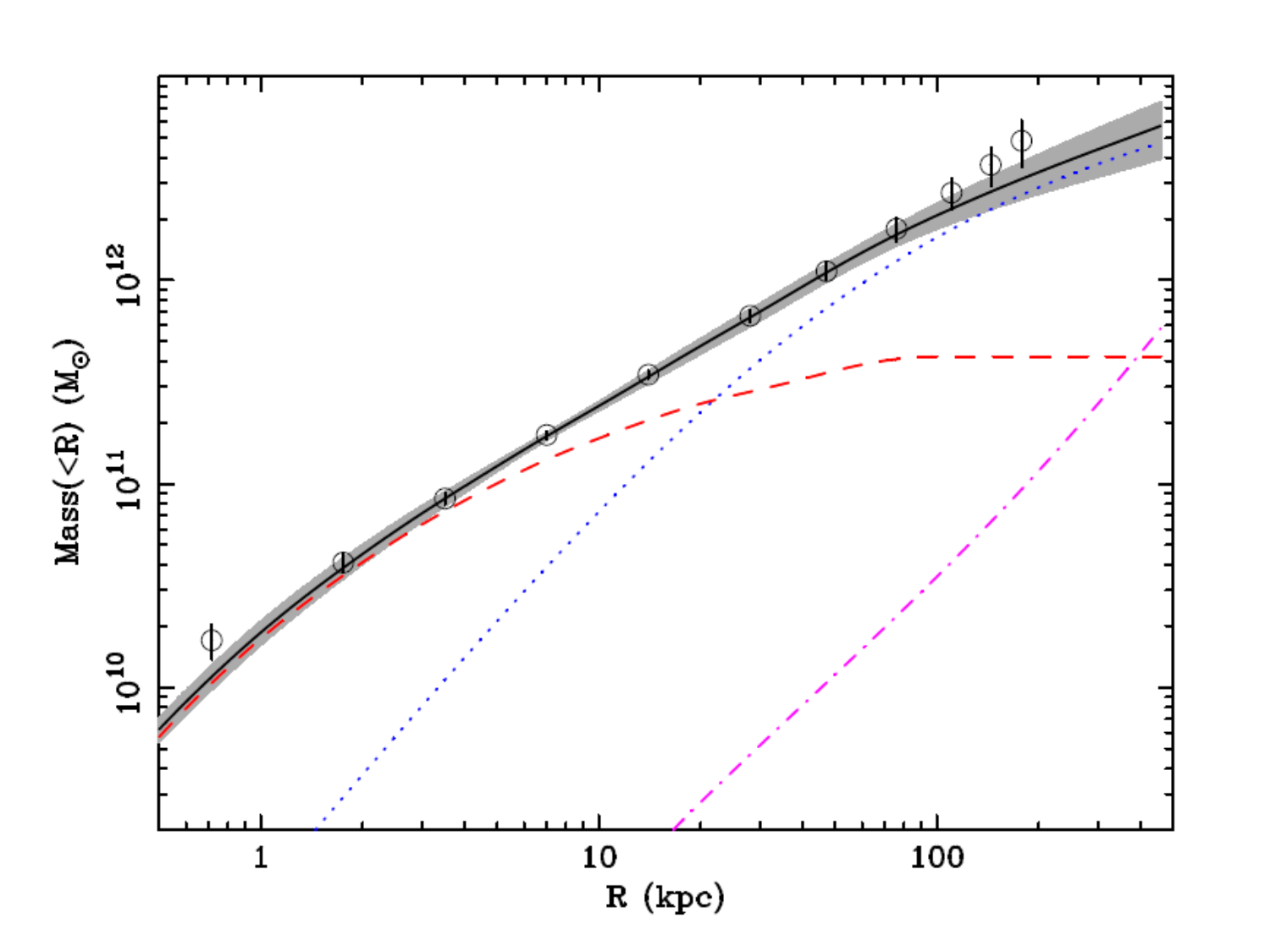}
   \caption{Radial mass profile of the elliptical galaxy NGC 1521 from a model calculation (not fitted to the measured points shown). The solid black line indicates the total enclosed mass ($1 \sigma$ errors in grey), the dashed red line is the stellar mass, the dotted blue line is the dark matter, and the dash-dot magenta line is the gas mass contribution. From P. J. Humphrey \& al. \cite{Humphrey}.}
 \end{figure}
Inside the half light radius $R_e$ the contribution of the dark matter halo to the central velocity dispersion is often very small, $<100$ km s$^{-1}$, so that the dark matter profile is intrinsically unresolvable. The outer mass profile is compatible with NFW, Eq.(\ref{halo}), and with Burkert, Eq.(\ref{Burkert}), as well. Important information on the mass distribution can be obtained from the Fundamental Plane, Eq.~(\ref{FundPlane}). which yields the coefficients $a=1.8$, $b=0.8$. Note that this is in some tension with the Virial Theorem, perhaps due to variations in the central dispersions, $\sigma_0$, of the stellar populations.

O. Tiret \& al. \cite{Tiret} concluded from a study of 23 giant elliptical galaxies with central velocity dispersions $\geq 330$ km s$^{-1}$, that the mass within $5-10$ kpc is dominated by the stars, not by DM. On the average the dark matter component contributes less than 5\% to the total velocity dispersions.

 The ELIXR survey is a volume-limited ($\leq 110$ Mpc) study by P. J. Humphrey \& al. \cite{Humphrey}, of optically selected, isolated, L* elliptical galaxies in particular the NGC 1521, for which X-ray data from \textit{Chandra} and \textit{XMM} exist. The isolation condition selects the appropriate galaxy halo and reduces the influence of a possible group-scale or cluster-scale halo.

 Most of the baryons are in a morphologically relaxed hot gas halo detectable out to $\approx 200$ kpc, that is well described by hydrostatic models.
 The baryons and the dark matter conspire to produce a total mass density profile that can be well-approximated by a power law, $\rho_{tot}\propto r^{-\alpha}$ over a wide range (as has been noted before, see references in \cite{Tiret, Humphrey}).

 The fitting method involves solving the equation of hydrostatic equilibrium to compute temperature and density profile models, given parametrized mass and entropy profiles. The models are then projected onto the sky and fitted to the projected temperature and density profiles. A fit ignoring DM was poor, but inclusion of DM improved the fit highly significantly: DM was required at $8.2 \sigma$. We show this fit in Fig. 9. In several studies \cite{Tiret, Salinas}, for most of the radii the dark matter contribution is very small although statistically significant.

\section{Mass to luminosity ratios and dwarf spheroidals}

The mass-to-light ratio of an astronomical object is defined as $\Upsilon\equiv M/L$. Stellar populations exhibit values $\Upsilon=1-10$ in solar units, in the solar neighborhood $\Upsilon =2.5-7$, in the Galactic disk $\Upsilon =1.0-1.7$ from C. Flynn \& $al.$ \cite{Flynn}.

Dwarf spheroidal galaxies (dSph) are the smallest stellar systems containing
dark matter and exhibit very high $M/L$ ratios, $\Upsilon =10-100$.
In Andromeda IX $\Upsilon$= 93 +120/-50, in Draco $\Upsilon=330\pm 125$.
The dwarf spheroidals have radii of $\approx 100$ pc and central velocity dispersions $\approx 10$ km s$^{-1}$ which is larger than expected for self-gravitating, equilibrium stellar populations. The generally accepted picture has been, that dwarf galaxies have slowly rising rotation curves and are dominated by dark matter at all radii.

However, R. A. Swaters \& al. \cite{Swaters} have reported observations of H I rotation curves for a sample of 73 dwarf galaxies, among which eight galaxies have sufficiently extended rotation curves to permit reliable determination of the core radius and the central density. They found that dark matter only becomes important at radii larger than three or four disk scale lengths. Their conclusion is, that the stellar disk
can explain the mass distribution over the optical parts of the
galaxy, and dark matter only becomes relevant at large radii. However,
the required stellar mass-to-light ratios are high, up to 15 in the R-band.

Comparing the properties of dwarf galaxies
in both the core and outskirts of the Perseus Cluster, Penny and Conselice \cite{Penny} found a clear correlation between mass-to-light ratio and the
luminosity of the dwarfs, such that the faintest dwarfs require
the largest fractions of dark matter to remain bound.
This is to be expected, as the fainter a galaxy is, the less luminous
mass it will contain, therefore the higher its dark
matter content must be to prevent its disruption.
Dwarfs are more easily influenced by their environment
than more massive galaxies

The distance to the Perseus Cluster prevents an easy determination of $\Upsilon$, so S. J. Penny \& C. J. Conselice \cite{Penny} instead determined the dark matter content
of the dwarfs by calculating the minimum mass needed in order to prevent tidal disruption by the cluster potential, using their sizes, the projected distance from the
cluster center to each dwarf and the mass of the cluster interior.
Three of 15 dwarfs turned out to have mass-to-light ratios smaller
than 3, indicating that they do not require dark matter.

Ultra-compact dwarf galaxies (UCDs) are stellar systems with masses of around $10^7 - 10^8$ M$_{sun}$ and half-mass radii of 10--100 pc. A remarkable properties of UCDs is
that their dynamical mass-to-light ratios are on average
about twice as large as those of globular clusters of
comparable metallicity, and also tend to be larger than
what one would expect based on simple stellar evolution
models. UCDs appear to contain very little or no dark matter.

H. Baumgardt \& S. Mieske \cite{Baumgardt} have presented collisional N-body simulations which study the coevolution of a system composed of stars and dark matter. They find that DM gets removed from the central regions of such systems due to dynamical friction
and mass segregation of stars. The friction timescale is significantly shorter than
a Hubble time for typical globular clusters, while most UCDs have friction times much
longer than a Hubble time. Therefore, a significant dark matter fraction may remain within
the half-mass radius of present-day UCDs, making dark matter a viable explanation
for their elevated mass-to-light ratios.

A different type of systems are the ultra-faint dwarf galaxies (UFDs). When interpreted as steady state objects in virial equilibrium by V. Belokurov \& $al.$ \cite{Belokurov}, they would be the most DM dominated objects known in the Universe. Their half-light radii range from 70 pc to 320 pc.

A special case is the UFD disk galaxy Segue 1, studied by M. Xiang-Gruess \& $al.$ \cite{Xiang}, which has a baryon mass of only about 1000 solar masses. One interpretation is that this is a thin non-rotating stellar disk not accompanied by a gas disk, embedded in an axisymmetric DM halo and with a ratio $f \equiv M_{halo}/M_b \approx 200$. But if the disk rotates, $f$ could be as high as 2000. If Segue 1 also has a magnetized gas disk, the dark matter halo has to confine the effective pressure in the stellar disk and the magnetic Lorentz force in the gas disk as well as possible rotation. Then $f$ could be very large \cite{Xiang}. Another interpretation is that Segue 1 is an extended globular cluster rather than an UFD \cite{Belokurov}.

\section{Small galaxy groups emitting X-rays}
There are examples of groups formed by a small number of galaxies
which are enveloped in a large cloud of hot gas (ICM), visible by its X-ray emission. One may assume that the electron density distribution associated with the X-ray brightness is in hydrostatic equilibrium, and one can extract the ICM radial density profiles by fits.

The amount of matter in the form of hot gas can be deduced from the intensity of this radiation. Adding the gas mass to the observed luminous matter, the total amount of baryonic matter, $M_b$, can be estimated, see M. Markevitch \& al. \cite{Markev99} and C. De Boni \& G. Bertin \cite{Boni}.
In clusters studied, the gas fraction increases with the distance from
the center; the dark matter appears more concentrated than the visible matter.

The temperature of the gas depends on the strength of the gravitational field, from which the total amount of gravitating matter, $M_{\rm grav}$, in the system can be deduced. In many such small galaxy groups one finds $M_{\rm grav}/M_b\geq3$, testifying to a dark halo present. An accurate estimate of $M_{\rm grav}$ requires that also dark energy is taken into account, because it reduces the strength of the gravitational potential. There are sometimes doubts whether all galaxies appearing near these groups are physical members. If not, they will artificially increase the velocity scatter and thus lead to larger virial masses.

On the scale of large clusters of galaxies like the Coma, it is generally observed that DM represents about 85\% of the total mass and that the visible matter is mostly in the form of a hot ICM.
   \begin{figure}[htbp]
   \includegraphics[width=15cm]{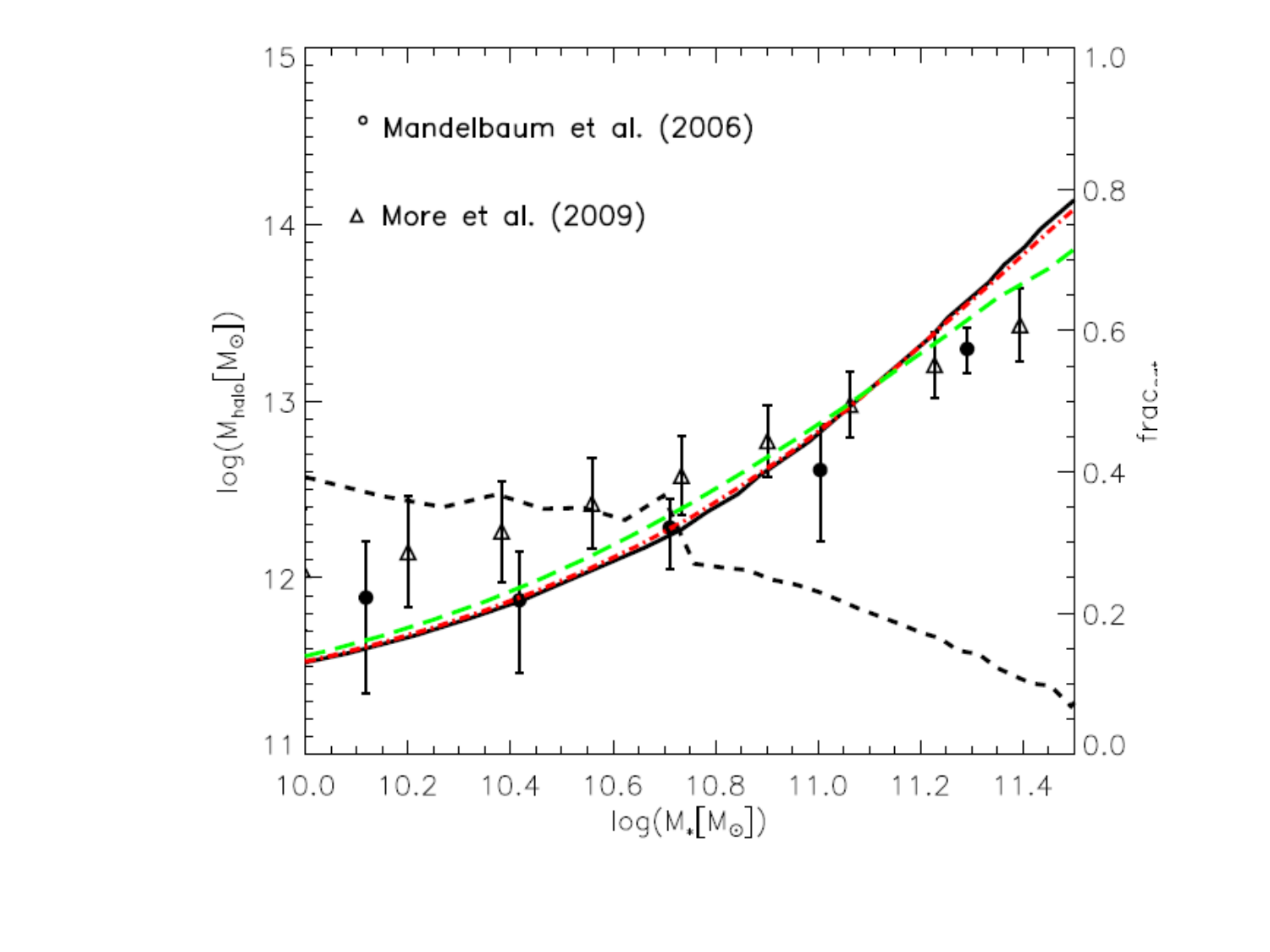}
   \caption{Dark matter halo mass $M_{halo}$ as a function of stellar mass $M_{\ast}$.
The thick black curve is the prediction from abundance matching
assuming no dispersion in the relation between the two masses. Red
and green dashed curves assume some dispersion in $\log M_{\ast}$. The dashed black curve is the satellite fraction as a function of stellar mass, as labeled on the axis at the right-hand side of the plot. From Qi Guo \& al. \cite{Guo}.}
\end{figure}
\section{Mass autocorrelation functions}
If galaxy formation is a local process, then on large scales galaxies must trace mass. This requires the study of how galaxies populate DM halos. In simulations one attempts to track galaxy and DM halo evolution across cosmic time in a physically consistent way, providing positions, velocities, star formation histories and other physical properties for the galaxy populations of interest.

Guo \& al. \cite{Guo} use abundance matching arguments to derive an accurate relation between galaxy stellar mass and DM halo mass. They combine a
stellar mass function based on spectroscopic observations with a precise halo/subhalo mass function obtained from simulations. Assuming this stellar mass - halo mass relation to be unique and monotonic, they compare it with direct observational estimates of the mean mass of halos surrounding galaxies of given stellar mass inferred from gravitational lensing and satellite galaxy dynamics data, and use it to populate halos in simulations. The stellar mass - halo mass relation is shown in Fig. 10.

      \begin{figure}[htbp]
   \includegraphics[width=12cm]{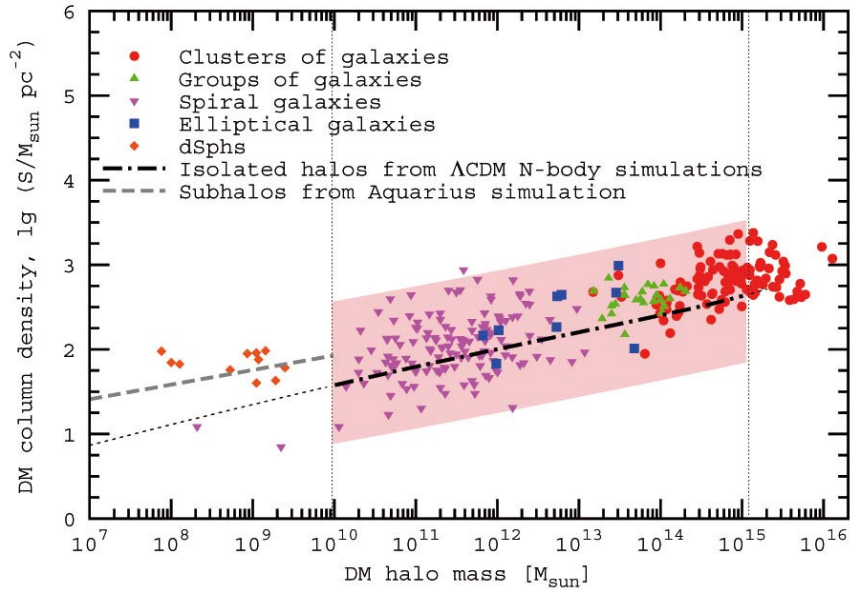}
   \caption{Dark matter column density vs. dark matter halo mass in solar units.  From A. Boyarsky \& {\em al.} \cite{Boyarsky}}
\end{figure}

The implied spatial clustering of stellar
mass turns out to be in remarkably good agreement with a direct and
precise measurement. By comparing the galaxy autocorrelation function with the total mass autocorrelation function, as averaged over the Local Supercluster (LSC) volume, one concludes that a large amount of matter in the LSC is dark.

A similar study is that of Boyarsky \& al. \cite{Boyarsky} who find a universal relation between DM column density and DM halo mass, satisfied by matter distributions at all observable scales in halo sizes from $10^8$ to $10^{16}$ M$_{sun}$, as shown in Fig. 11. Such a universal property is difficult to explain without dark matter.

\section{Cosmic Microwave Background (CMB)}

\voffset= 0.4cm
\begin{figure}[htbp]
   \includegraphics[width=15.0cm]{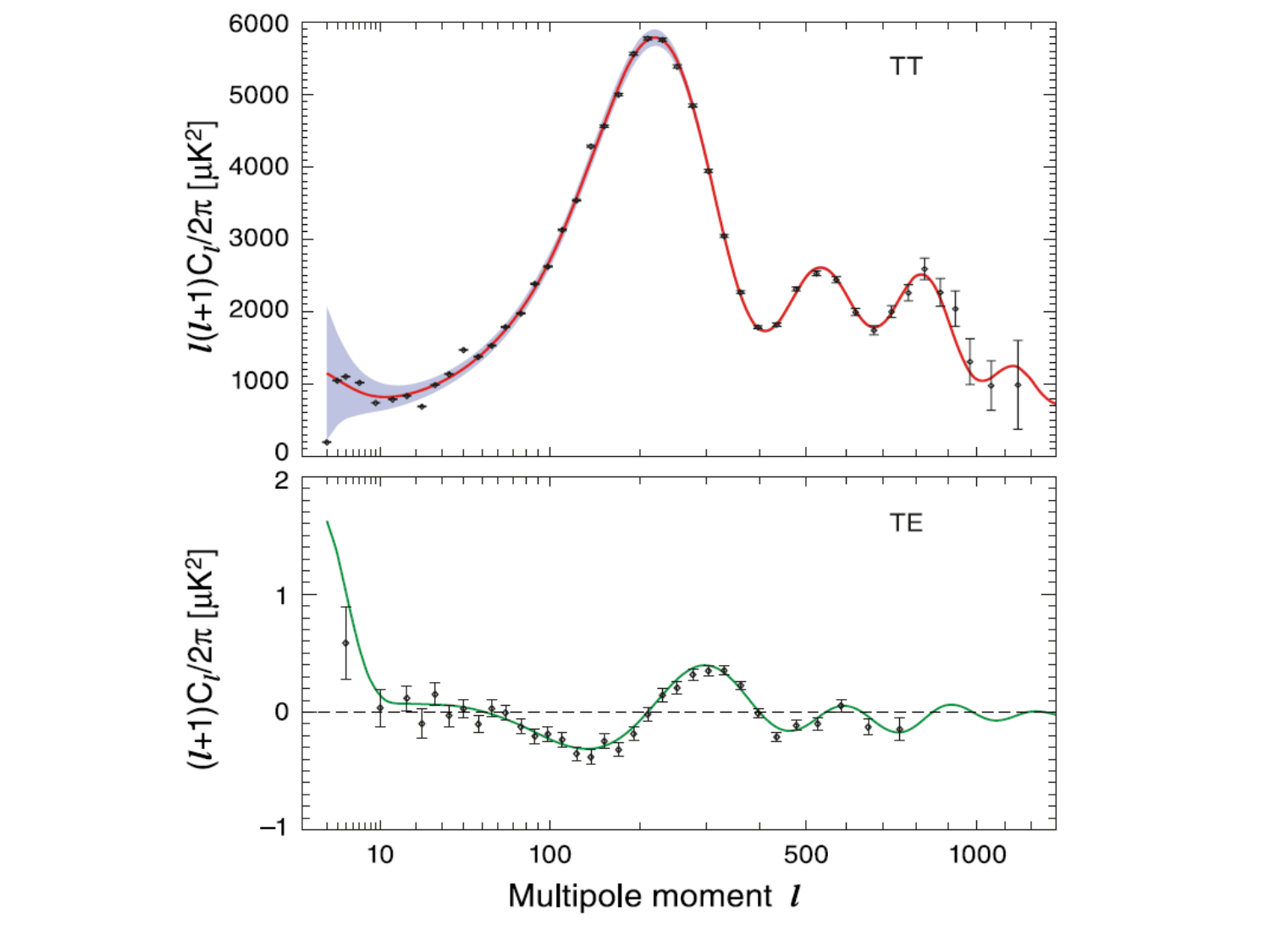}
   \caption{The CMB radiation temperature (TT) and temperature-polarization (TE) power spectra from the seven-year WMAP 94 GHz maps.show anisotropies which can be analyzed by power spectra as functions of multipole moments. The solid line shows the best-fit prediction for the flat $\Lambda$CDM model. From D. Larson \& {\em al.} \cite{Larson}}.
      \end{figure}

The tight coupling between radiation and matter density before decoupling
caused the primordial adiabatic perturbations to oscillate in phase.
Beginning from the time of last scattering, the receding horizon has been revealing these
frozen density perturbations, setting up a pattern of standing
acoustic waves in the baryon-photon fluid.  After
decoupling, this pattern is visible today as temperature anisotropies with a certain regularity across the sky.

The primordial photons are polarized by the anisotropic Thomson
scattering process, but as long as the photons continue to meet free
electrons their polarization is washed out, and no net polarization
is produced. At a photon's last scattering however, the induced
polarization remains and the subsequently free-streaming photon
possesses a quadrupole moment.

Temperature and polarization fluctuations are analyzed in terms of multipole components or powers. The resulting distribution of powers versus multipole $\ell$, or multipole moment $k=2\pi/\ell$, is the \textit{power spectrum} which exhibits conspicuous \emph{Doppler peaks}. In Fig. 12 we display the radiation temperature (TT) and temperature- E-polarization correlation (TE) power spectra from the 7-year data of WMAP as functions of multipole moments \cite{Larson}. The spectra can then be compared to theory, and theoretical parameters determined. Many experiments have determined the power spectra, so a wealth of data exists.

Baryonic matter feels attractive self-gravity and is pressure-supported, whereas dark matter only feels attractive self-gravity, but is pressureless. Thus the Doppler peaks in the CMBR power spectrum testify about baryonic and DM, whereas the troughs testify about rarefaction caused by the baryonic pressure. The position of the first peak determines $\Omega_m h^2$. Combining the TT data with determinations of the Hubble constant $h$, the WMAP team can determine the total mass density parameter $\Omega_m=\Omega_b+\Omega_{\rm dm}$. The ratio of amplitudes of the second-to-first Doppler peaks determines the baryonic density parameter to be $\Omega_b= 0.0449\pm 0.0028$ and the dark matter component to be $\Omega_{\rm dm}= 0.222\pm 0.026$ \cite{Larson}, thus $\Omega_m=0.267$.
\begin{figure}[htbp]
   \includegraphics[width=15cm]{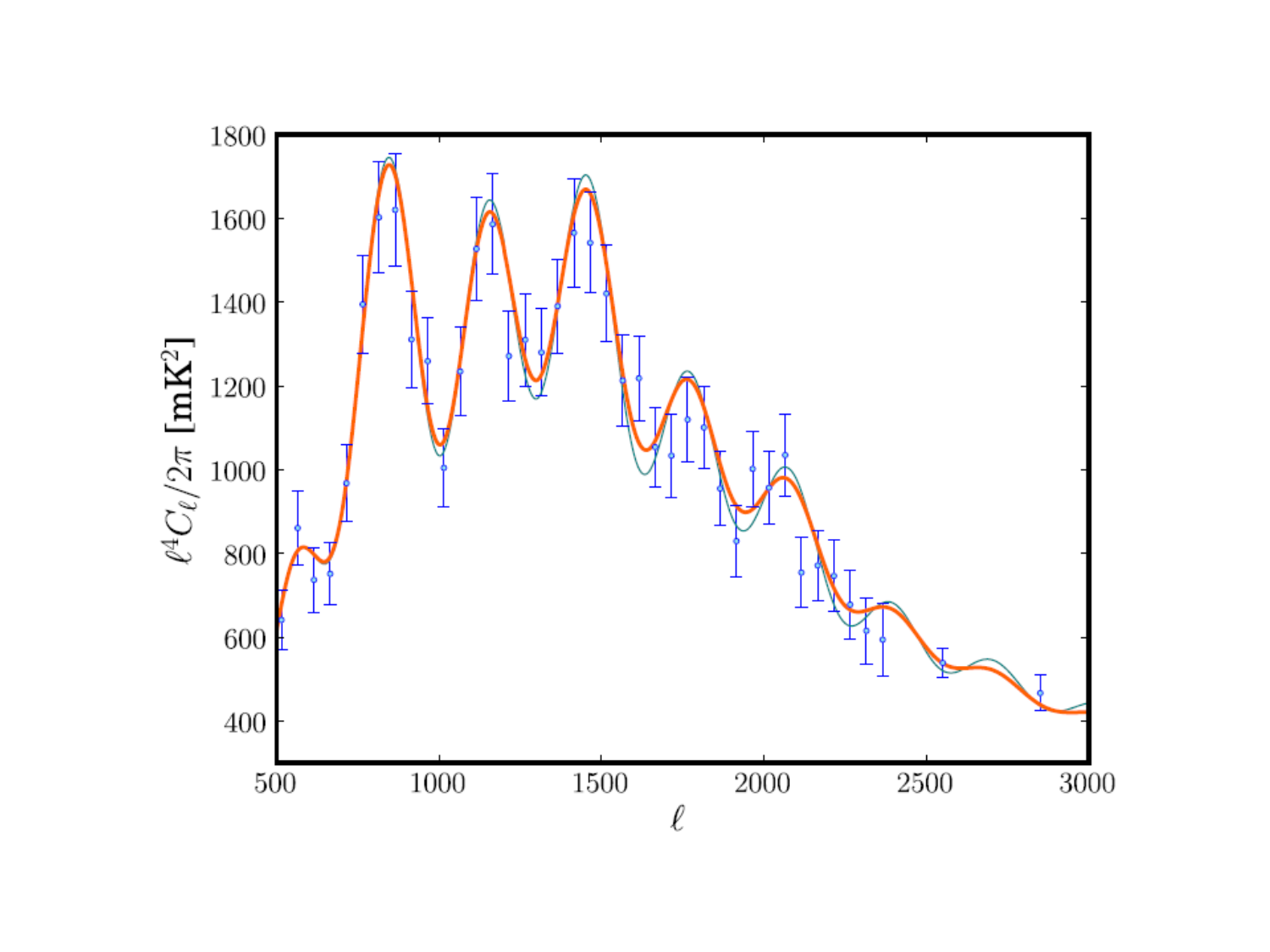}
   \caption{The ACT 148GHz power spectrum multiplied by $\ell^4$ is shown for lensed (orange curve) and unlensed models (green curve). From S. Das \& $al.$ \cite{Das}.}
      \end{figure}

Power spectra at higher multipole moments have been measured with the Atacama Cosmology Telescope (ACT) \cite{Das} at 148 GHz and 218 GHz, as well as the cross-frequency spectrum
between these two channels. and found to be in agreement with the 7-year WMAP 94 GHz maps in the common range $400\leq\ell\leq 1000$. The ACT has also been able to measure the lensing of the CMB signal at a significance of $2.8 \sigma$, which slightly smooths out the acoustic peaks, \textit{cf} Fig. 13.

In a fit of the flat $\Lambda$CDM model to the data the dark matter density parameter comes out slightly higher than WMAP and the baryonic density slightly lower so the total density parameter for WMAP and ACT added is $\Omega_m=0.276\pm 0.016$ \cite{Dunkley}.
\begin{figure}[htbp]
   \includegraphics[width=15cm]{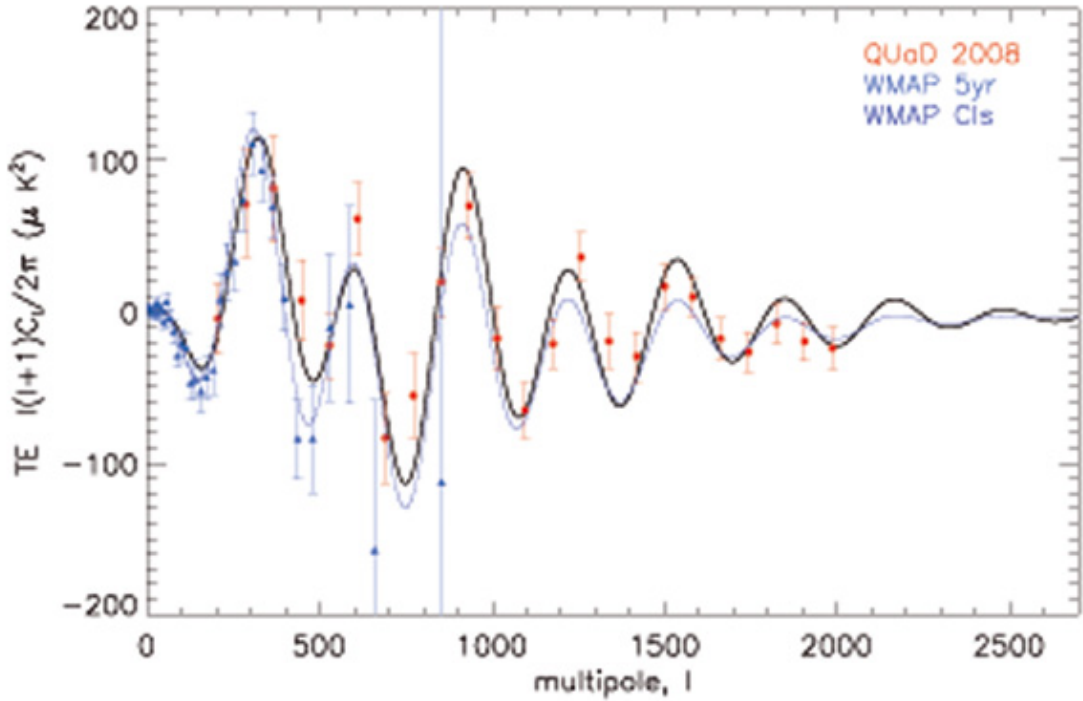}
   \caption{The E-mode polarization power spectrum (EE) from the CMB observations of the QUaD collaboration \& {\em al.} \cite{QUAD}}
\end{figure}
Information on the TE correlations comes from several measurements, among them WMAP \cite{Larson}, and on the E-mode polarization power spectrum alone (EE) from the QUAD collaboration \cite{QUAD}, Fig. 14.

The results show two surprises: Firstly, since $\Omega_m\ll 1$, a large component $\Omega_{\Lambda}\approx 0.74$ is missing,
of unknown nature, and termed \textit{dark energy}. The second
surprise is that ordinary baryonic matter is only a small fraction
of the total matter budget. The remainder is then dark matter, of unknown composition.
Of the 4.5\% of baryons in the Universe only about 1\% is stars.

\section{Baryonic acoustic oscillations (BAO)}

A cornerstone of cosmology is the Copernican principle, that matter in the Universe is distributed homogeneously, if only on the largest scales of superclusters separated by voids. On smaller scales we observe inhomogeneities in the forms of galaxies, galaxy groups, and clusters. The common approach to this situation is to turn to non-relativistic hydrodynamics and treat matter in the Universe as an adiabatic, viscous, non-static fluid, in which random fluctuations around the mean density appear, manifested by compressions in some regions and rarefactions in other. The origin of these density fluctuations was the tight coupling
established before decoupling between radiation and charged matter density, causing them to oscillate in phase. An ordinary fluid is dominated by the material pressure, but in the fluid of our Universe three effects are competing: gravitational attraction, density dilution due to the Hubble flow, and radiation pressure felt by charged particles only.

The inflationary fluctuations crossed the post-inflationary Hubble radius, to come back into vision with a wavelength corresponding to the size of
the Hubble radius at that moment. At time $t_{eq}$ the overdensities began to amplify and grow into larger inhomogeneities. In overdense regions where the gravitational forces dominate, matter contracts locally and attracts surrounding matter, becoming increasingly unstable until it eventually collapses into a gravitationally bound object. In regions where the pressure forces dominate, the fluctuations move with constant amplitude as sound waves in the fluid, transporting energy from one region of space to another.

\begin{figure}[htbp]
   \includegraphics[width=12cm]{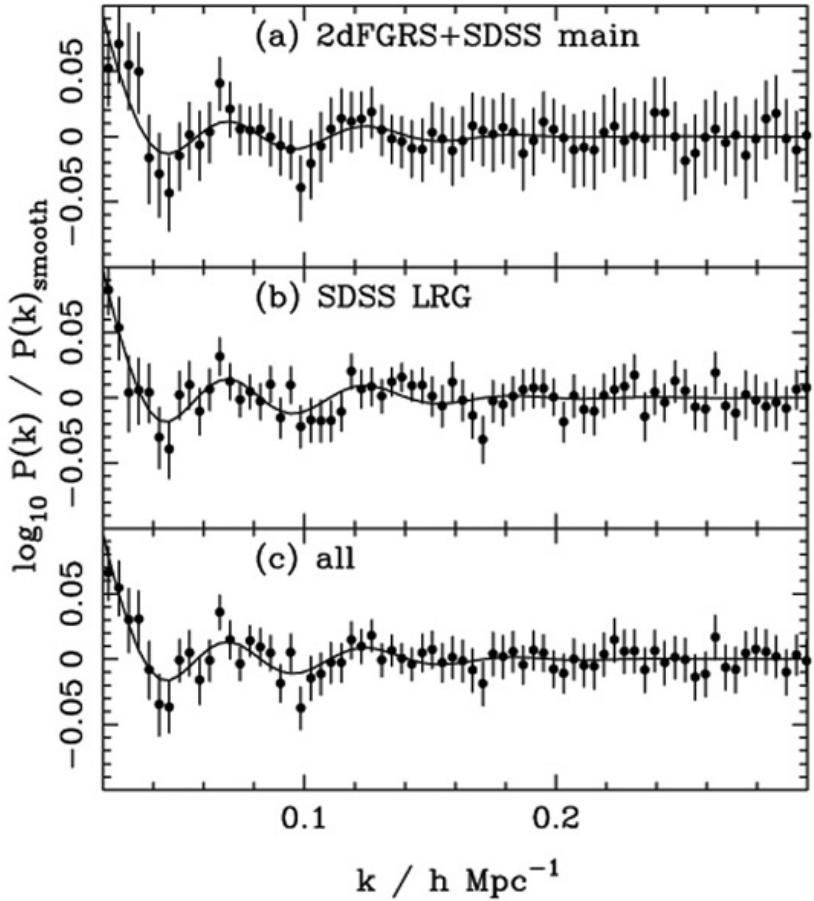}
   \caption{BAO in power spectra calculated from (a) the combined SDSS
and 2dFGRS main galaxies, (b) the SDSS DR5 LRG sample, and (c) the
combination of these two samples (solid symbols with $1\sigma$ errors). The data
are correlated and the errors are calculated from the diagonal terms in the
covariance matrix. A Standard $\Lambda$CDM distance--redshift relation was assumed
to calculate the power spectra with $\Omega_m = 0.25,~\Omega_{\Lambda} = 0.75$. From W. J. Percival \& {\em al.} \cite{BAO}.}
      \end{figure}

Inflationary models predict that the primordial mass density fluctuations should be adiabatic, Gaussian, and exhibit the same scale invariance as the CMB fluctuations.
The baryonic acoustic oscillations can be treated similarly to CMB, they are specified by the dimensionless \textit{mass autocorrelation function} which is the Fourier transform of the power spectrum of a spherical harmonic expansion. The power spectrum is shown in Fig. 15 \cite{BAO}.

As the Universe approached decoupling, the photon mean free path increased
and radiation could diffuse from overdense regions into underdense ones, thereby smoothing out any inhomogeneities in the plasma. The situation changed dramatically at recombination, at time 380\,000~yr after Big Bang, when all the free electrons suddenly disappeared, captured into atomic Bohr orbits, and the radiation pressure almost vanished. Now the baryon acoustic waves and the CMB continued to oscillate independently, but adiabatically, and the density perturbations which had entered the Hubble radius since then could grow with full vigor into baryonic structures.

The scale of BAO depends on $\Omega_m$ and on the Hubble constant, $h$, so one needs information on $h$ to break the degeneracy. The result is then $\Omega_m\approx0.26$. In the ratio $\Omega_b/\Omega_m$ the $h$-dependence cancels out, so one can also quantify the amount of DM on very large scales by $\Omega_b/\Omega_m=0.18\pm 0.04$.

\section{Galaxy formation in purely baryonic matter?}
We have seen in Sec. 10 that the baryonic density parameter, $\Omega_b$, is very small. The critical density $\Omega_{crit}$ is determined by the expansion speed of the Universe, and the mean baryonic density of the Universe (stars, interstellar and intergalactic gas) is only $\Omega_b = 0.045$ \cite{Larson}.

The question arises whether the galaxies could have formed from primordial density fluctuations in a purely baryonic medium. We have also noted, that the fluctuations in CMB and BAO maintain adiabaticity. The amplitude of the primordial baryon density fluctuations would have needed to be very large in order to form the observed number of galaxies. But then the amplitude of the CMB fluctuations would also have been very large, leading to intolerably large CMB anisotropies today. Thus galaxy formation in purely baryonic matter is ruled out by this argument alone.

Thus one concludes, that the galaxies could only have been formed in the presence of gravitating dark matter which started to fluctuate early, unhindered by radiation pressure. This conclusion is further strengthened in the next Section.

\section{Large Scale Structures simulated}
In the $\Lambda$CDM paradigm, the
nonlinear growth of DM structure is a well-posed
problem where both the initial conditions and the evolution
equations are known (at least when the effects of the baryons can be neglected).

The Aquarius Project \cite{Aquarius} is a Virgo Consortium program to carry out
high-resolution DM simulations of Milky-Way--sized
halos in the $\Lambda$CDM cosmology. This project seeks clues to
the formation of galaxies and to the nature of the dark matter
by designing strategies for exploring the formation of our
Galaxy and its luminous and dark satellites.
\begin{figure}[htbp]
   \includegraphics[width=15cm]{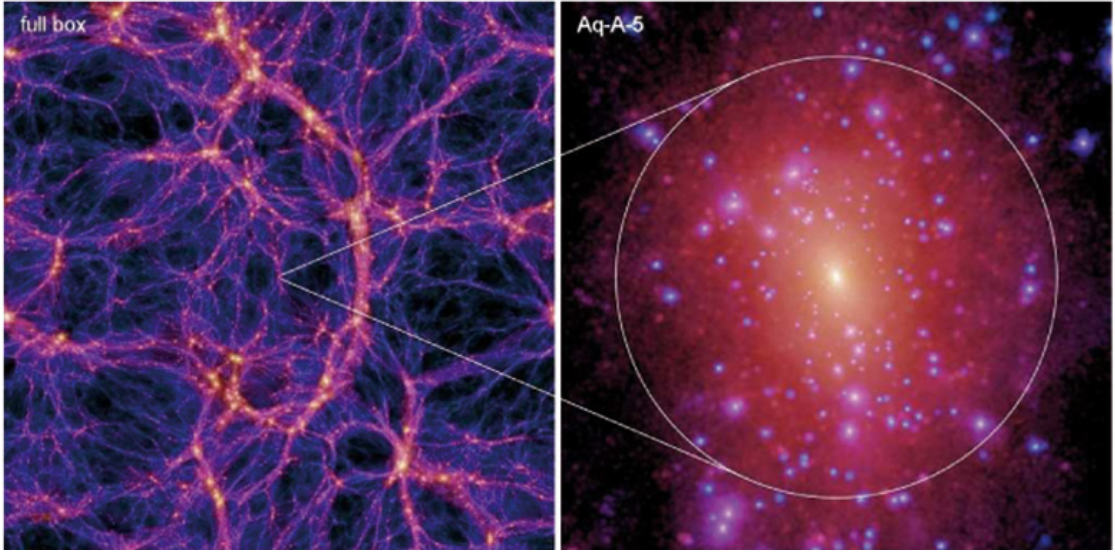}
   \caption{The left panel shows the projected dark matter density at z = 0 in a slice of thickness 13.7 Mpc through the full box
(137 Mpc on a side) of the $900^3$ parent simulation. The right panel
show this halo resimulated at a different numerical resolution. The image brightness is proportional to the logarithm of the squared DM density
projected along the line-of-sight. The circles mark $r_{50}$, the radius within which the mean density is 50 times the background density. From V. Springel \& al. \cite{Aquarius}}
\end{figure}
The galaxy population on scales from 50 kpc to the size of the observable Universe has been predicted by hierarchical $\Lambda$CDM scenarios, and compared directly with a wide array of observations. So far, the $\Lambda$CDM paradigm has passed these tests successfully, particularly those that consider the large-scale matter distribution and has led to the discovery of a universal internal structure for DM halos.
As was noted in Sec. 12, the observed structure of galaxies, clusters and superclusters, as illustrated by Fig. 16, could not have formed in a baryonic medium devoid of dark matter.

Given this success, it is important to test $\Lambda$CDM predictions
also on smaller scales, not least because these are sensitive
to the nature of the dark matter. Indeed, a number of serious
challenges to the paradigm have emerged on the scale
of individual galaxies and their central structure. In particular,
the abundance of small DM subhalos predicted
within CDM halos is much larger than the number of known satellite galaxies
surrounding the Milky Way (M. Boylan-Konchin \& $al.$ \cite{Boylan} and references therein).

\section{Dark matter from overall fits}
In Sec. 10 we have seen that the WMAP 7-year CMB data together with the Hubble constant value testify about the existence of DM \cite{Larson, Dunkley}. In Sec. 11 we addressed the BAO data \cite{BAO} with the same conclusion. In overall fits one combines these with supernova data (SN Ia) which offer a constraint nearly orthogonal to that of CMB in the $\Omega_{\Lambda}-\Omega_m$-plane.
\begin{figure}[htbp]
   \includegraphics[width=13cm]{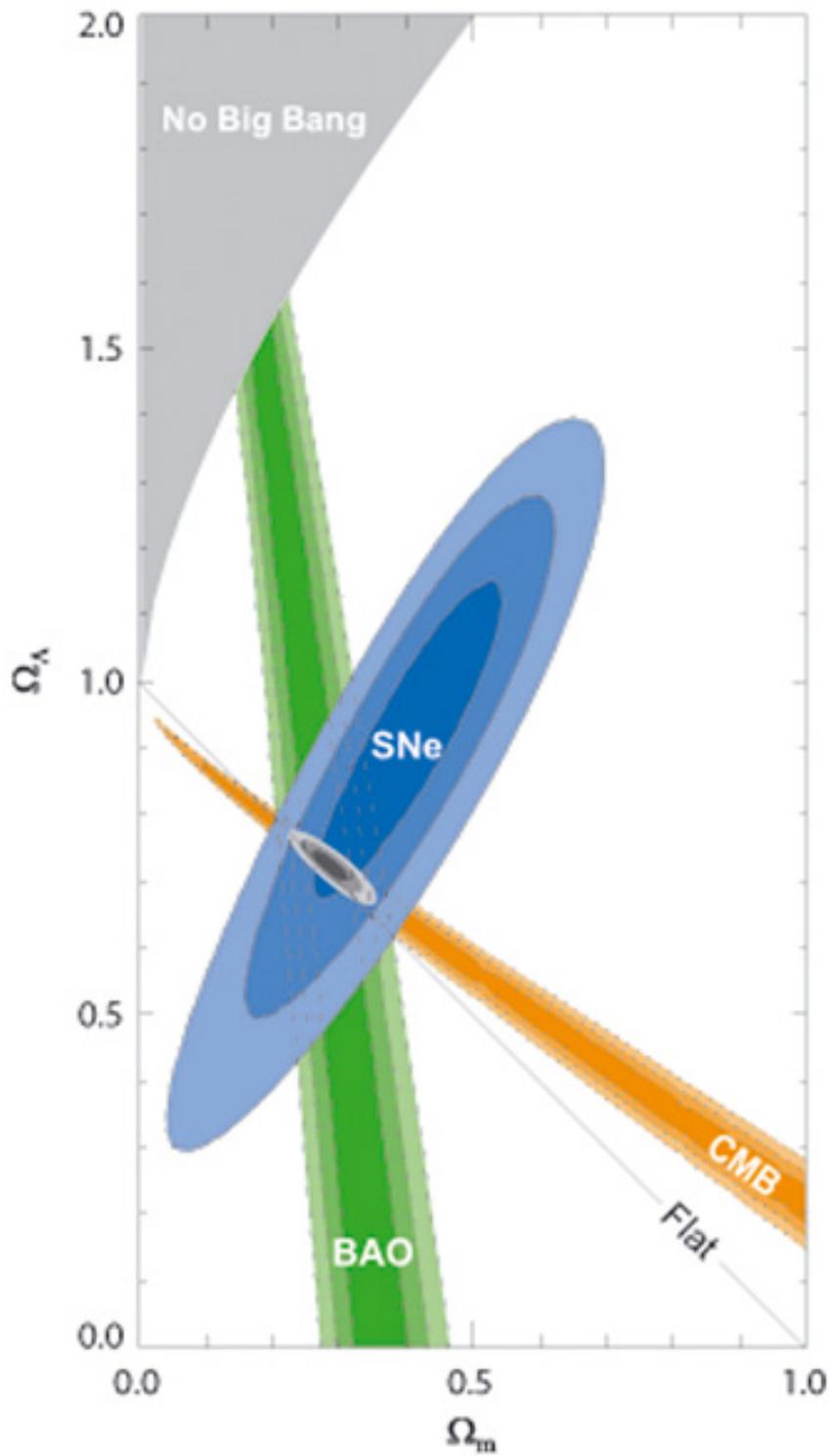}
   \caption{68.3 \%, 95.4 \% and 99.7\% confidence level contours on $\Omega_{\Lambda}$ and $\Omega_m$ obtained from CMB, BAO and the Union SN set, as well as their combination (assuming w = -1). Note the straight line corresponding to a flat Universe with $\Omega_{\Lambda}+\Omega_m=1$. From M. Kowalski \& al. \cite{Kowalski}.}
\end{figure}
The �Union� compilation of 307 selected SN Ia includes the recent large samples of SNe Ia from the Supernova Legacy Survey, the ESSENCE Survey, the older data sets, as well as the recently extended data set of distant supernovae observed with HST. M. Kowalski \& al. \cite{Kowalski} present the latest results from this compilation and discuss the cosmological constraints and its combination with CMB and BAO measurements. The CMB constraint is close to the line $\Omega_{\Lambda}+\Omega_m=1$, whereas the supernova constraint is close to the line $\Omega_{\Lambda}-1.6\times\Omega_m=0.2$. The BAO data constrain $\Omega_m$, but hardly at all $\Omega_{\Lambda}$. This is shown in Fig. 17.

Defining the vacuum energy density parameter by $\Omega_k=1-\Omega_{\Lambda}-\Omega_m$,
a flat Universe corresponds to $\Omega_k=0$. For a non-flat $\Lambda CDM$ Universe with a cosmological constant responsible for dark energy, a simultaneous fit to the data sets gives
\begin{equation}
\Omega_m=0.285+0.020/-0.019\pm 0.011,\\
\Omega_k=-0.009+0.009+0.002/-0.010-0.003\ ,
\end{equation}
where the first error is statistical and the second error systematic. Clearly one notes that the Universe is consistent with being flat. Subtracting $\Omega_b = 0.045$ from $\Omega_m=0.285$ one obtains the density parameter for DM, $\Omega_{dm}\approx 0.24$. Assuming flatness, M. Kowalski \& al. \cite{Kowalski} find $\Omega_m=0.274\pm 0.016\pm 0.013$. This compares well with the combined 7-year WMAP data and the ACT data, $\Omega_m=0.276\pm 0.016$ \cite{Dunkley}. If one fits different models having more free parameters, one gets slightly different results, but all within these $1\sigma$ errors.

\begin{figure}[htbp]
   \includegraphics[width=13cm]{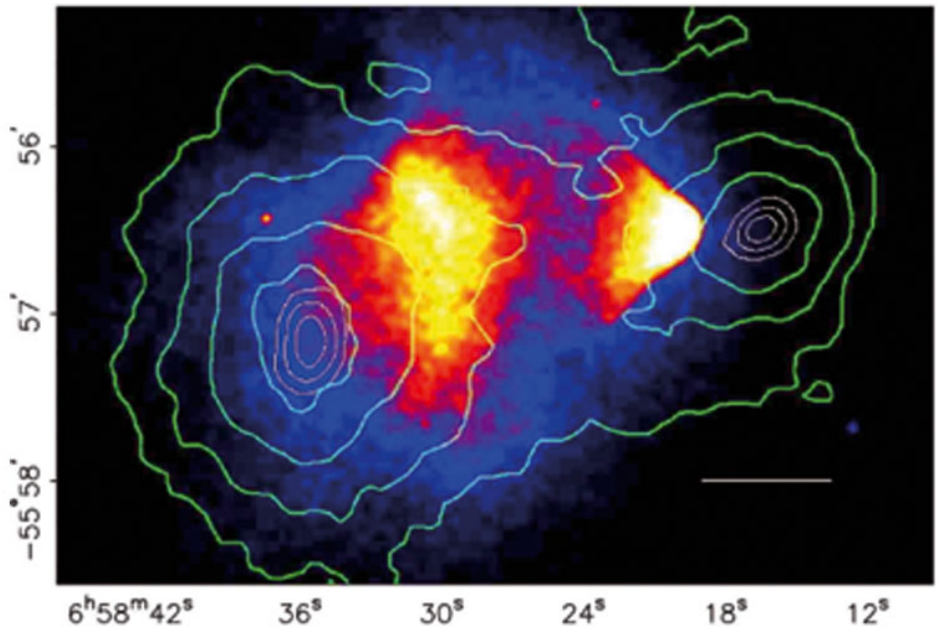}
   \caption{The merging cluster 1E0657-558. On the right is the smaller \textit{Bullet cluster} which has traversed the larger cluster. The colors indicate the X-ray temperature of the plasma: blue is coolest and white is hottest. The green contours are the weak lensing reconstruction of the gravitational potential of the cluster. From D. Clowe \& al. \cite{Clowe}}
\end{figure}

\section{Merging galaxy clusters}
In isolated galaxies and galaxy clusters all matter components contributing to the common gravitational potential are more or less centrally-symmetrically coincident. This makes the dissection of DM from the baryonic components difficult and dependent on parametrization, as we have discussed in Sec. 3. In merging galaxy clusters however, the separate distributions of galaxies, intracluster gas and DM may become spatially segregated permitting separate observations. The visually observable galaxies behave as collisionless particles, the baryonic intracluster plasma is fluid-like, experiences ram pressure and emits X-rays, but non-interacting DM does not feel that pressure, it only makes itself felt by its contribution to the common gravitational potential.

Major cluster mergers are the most energetic events in the Universe since the Big Bang. Shock fronts in the intracluster gas are the key observational tools in the study of these systems. When a subcluster traverses a larger cluster it cannot be treated as a solid body with constant mass moving at constant velocity. During its passage through the gravitational potential of the main cluster it is shrinking over time, stripped of gas envelope and decelerating. Depending on the ratio of the cluster masses, the gas forms a bow shock in front of the main cluster, and this can even be reversed at the time when the potentials coincide.

 We shall now meet several examples of galaxy cluster mergers where the presence of DM could be inferred from the separation of the gravitational potential from the position of the radiating plasma.
\subsection{The Bullet cluster 1E0657-558}
The exceptionally hot and X-ray luminous galaxy cluster 1E0657-558, the \textit{Bullet cluster} at redshift $z=0.296$, was discovered by Tucker et al. in 1995 \cite{Tucker} in \textit{Chandra} X-ray data. Its structure as a merger of a 2.3 $\times 10^{14}$ M$_{sun}$ subcluster with a main 2.8 $\times 10^{14}$ M$_{sun}$ cluster was demonstrated by Markevitch et al. \cite{Markevitch, Markev04} and Clowe et al. \cite{Clowe04, Clowe}. This was presented as the first clear example of a bow shock in a heated intracluster plasma.

With the advent of high-resolution lensing Brada$\check{c}$ et al. \cite{Bradac05, Bradac06} developed a technique combining multiple strongly-lensed \textit{Hubble Space Telescope} multi-color images of identified galaxies, with weakly lensed and elliptically distorted background sources. The reconstructed gravitational potential does not trace the X-ray plasma distribution which is the dominant baryonic mass component, but rather approximately traces the distribution of bright cluster member galaxies, \textit{cf} Fig. 18.

 The center of the total mass is offset from the center of the baryonic mass peaks, proving that the majority of the matter in the system is unseen. In front of the bullet cluster which has traversed the larger one about 100 Myr ago with a relative velocity of 4500 km s$^{-1}$, a bow shock is evident in the X-rays. The main cluster peak and the distinct subcluster mass concentration are both clearly offset from the location of the X-ray gas \cite{Bradac06}. A recent analysis of this system \cite{Paraficz} confirms the results of references \cite{Clowe, Clowe04, Bradac06}, and in addition finds that dark matter forms three distinct clumps.

\begin{figure}[htbp]
   \includegraphics[width=13cm]{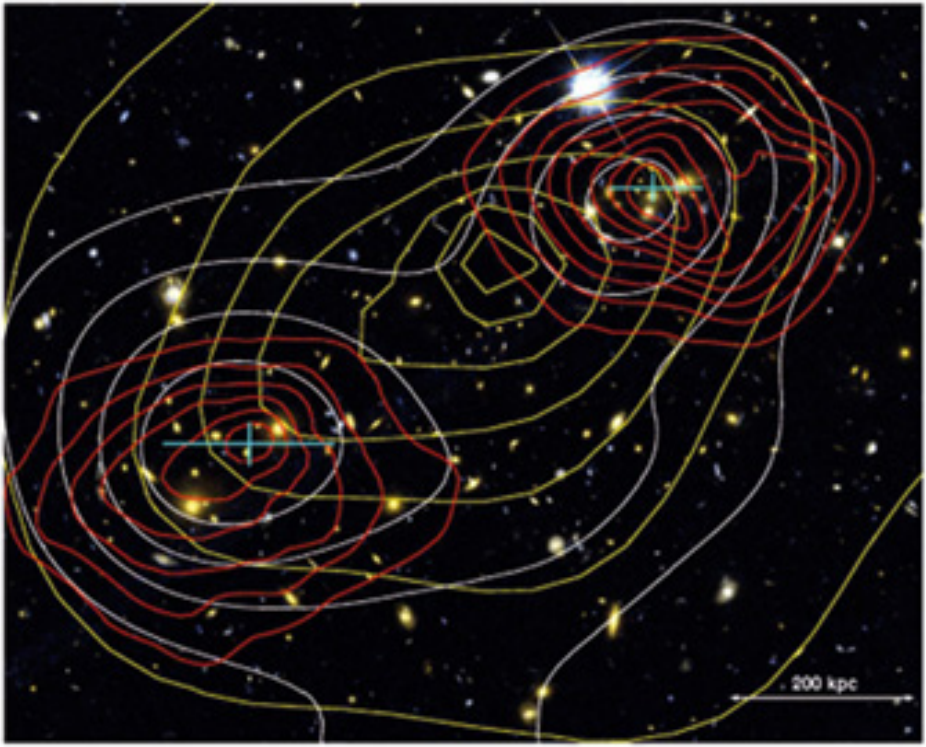}
   \caption{The color composite of the cluster MACS J0025.4-1222. Overlaid in red contours is the surface mass density (linearly spaced) from the
combined weak and strong lensing mass reconstruction. The X-ray brightness contours (also linearly spaced) are overlaid in yellow and the I-band light is
overlaid in white. The measured peak positions and error bars for the total mass of the two cluster components are shown as cyan crosses. From M. Brada$\check{c}$ \& $al.$ \cite{Bradac}.}
\end{figure}

\begin{figure}[htbp]
   \includegraphics[width=13cm]{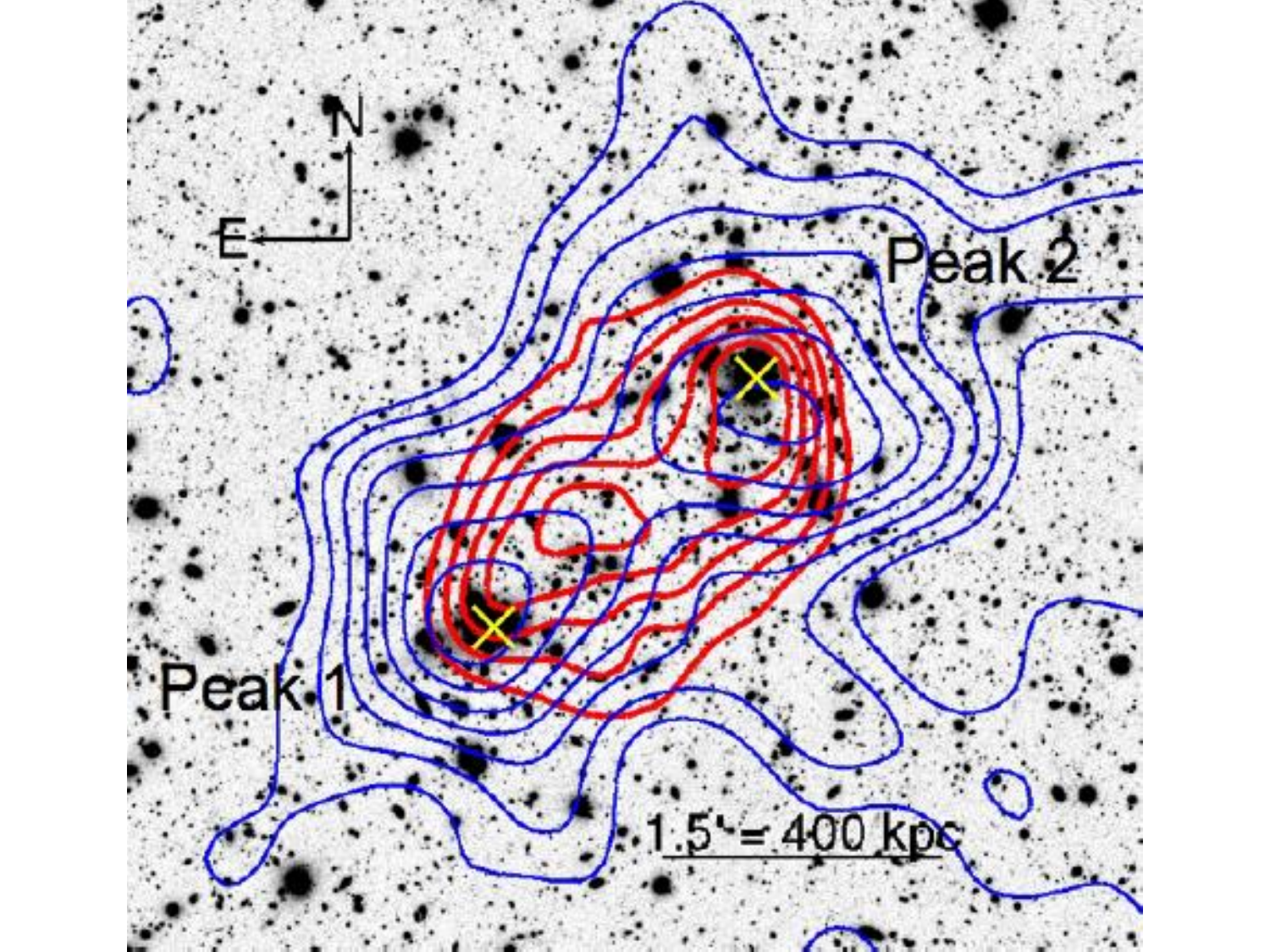}
   \caption{The A1758N merger from B. Ragozzine \& al. \cite{Ragozzine}. The blue contours represent the weak lensing mass reconstruction made from a background galaxy density of 24.0 galaxies/arcmin$^2$. The outer blue contour begins at surface mass density $\kappa=0.07$ and each contour increases in steps of 0.045 up to $\kappa=0.34$. The red contours follow the X-ray gas mass obtained in the Chandra exposure. The NW cluster's BCG aligns with the X-ray gas and the weak lensing peak. The SE cluster's BCG and weak lensing peak are well separated from the X-ray gas, which has a bright peak near the midpoint of the two weak lensing peaks.}
\end{figure}

\subsection{The galaxy cluster pair MACS J0025.4-1222}

Another merging system with similar characteristics but with lower spatial resolution has been reported by Brada$\check{c}$ et al. \cite{Bradac}, the post-merging galaxy cluster pair MACS J0025.4-1222, also called the \textit{Baby Bullet}. It has an apparently simple geometry, consisting of two large subclusters of similar richness, about 2.5 $\times 10^{14}$ M$_{sun}$, both at redshift $z=0.586$, colliding in approximately the plane of the sky. Multiple images due to strong lensing of four distinct components could be identified. The combined strong and weak lensing analysis follows the method in ref. \cite{Bradac06}.

The two distinct mass peaks are clearly offset by $4\sigma$ from the main baryonic component, which is the radiating hot gas observed by \textit{Chandra}. The relative merging velocity is estimated to be 2000 km s$^{-1}$. In Fig.19 we show linearly spaced surface mass density contours and X-ray brightness contours. The majority of the mass is spatially coincident with the identified galaxies which implies, that the cluster must be dominated by a relatively collisionless form of dark matter.

\subsection{The merging system A1758}
A much more complicated merging system is A1758 at redshift $z=0.279$, analyzed by the same team as above, B. Ragozzine \& al. \cite{Ragozzine}, and consisting of four clusters undergoing two separate mergers.  The weak lensing mass peaks of the two northern clusters A1758N are separated at the 2.5 $\sigma$ level, whereas the two southern clusters are not well separated and have a disturbed X-ray morphology. There is no evidence for a merger between A1758N and A1758S in the X-ray signature and they have a projected separation of 2.0 Mpc. Note however the SZ results from the Arcminute
Microkelvin Imager (AMI) in Cambridge (UK) on this system  \cite{Carmen, AMI},
which sees a hint of a signal between the A1758N and A1758S.

A1758N introduces a new geometry that is different from the previously discussed mergers: one weak lensing peak overlaps an X-ray peak, while the other weak lensing peak is clearly separated from the X-ray component, \textit{cf} Fig. 20.

Since no strong lensing has yet been confirmed, conclusions about cluster masses and DM would have to wait for better lensing data.

\subsection{The merging cluster Abell 2146}
\voffset= -2.0cm
\begin{figure}[htbp]
   \includegraphics[width=15cm]{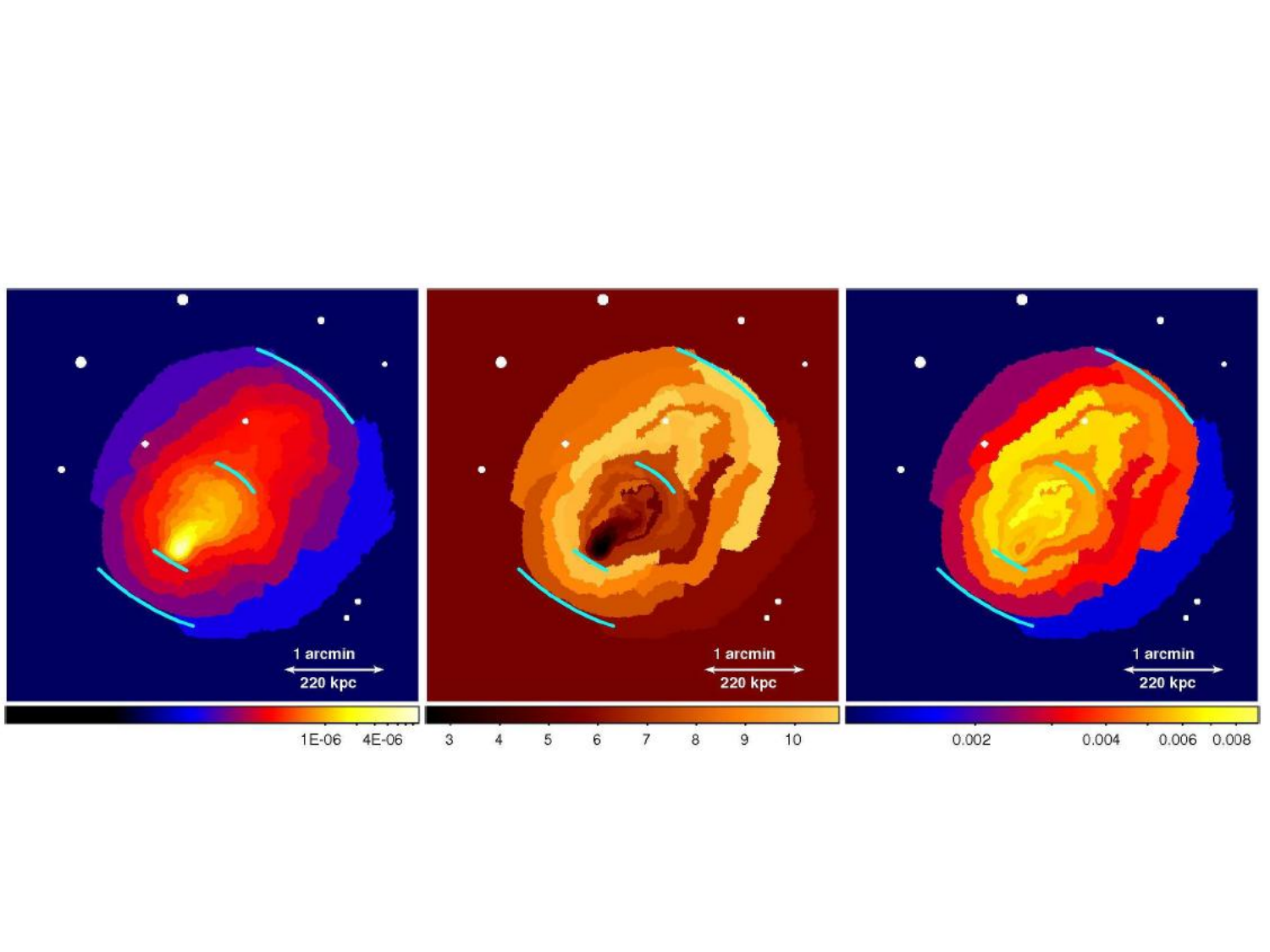}
   \caption{Left: Projected emission measure per unit area map for the merging cluster Abell 2146. Center: Projected temperature map. Right: Projected "pressure" map produced by multiplying the emission measure and temperature maps. From H. R. Russell \& $al$. \cite{Russell}}
\end{figure}

\textit{Chandra} observations of the cluster Abell 2146 at a redshift of $z=0.234$ have revealed two shock fronts, H. R. Russell \& $al.$ \cite{Russell}. The X-ray morphology suggests a recent merger where a subcluster containing a dense core has passed through the center of a second cluster, the remnant of which appears as the concentration of gas to the NW. The strongly peaked core has just emerged from the primary core, and is trailing material that has been ram pressure stripped in the gravitational potential. This material appears as a warmer stream of gas behind the subcluster core, and trails back to the hottest region of the disrupted main cluster.

Four steep surface brightness edges can be defined: two in the SE sector in front of the subcluster core and another two in the NW sector, \textit{cf}. Fig. 21. The interpretation is \cite{Russell} that an upstream shock is generated as the gravitational potential minimum fluctuates rapidly during the core passage, reaching an extreme minimum when the two cluster cores coalesce. This causes a significant amount of the outer cluster gas to flow inwards. When the subcluster core exits the main core the gravitational potential rapidly returns to its premerger level and expels much of the newly arrived gas which in turn collides with the residual infall, forming an inward traveling shock front. Behind the subcluster, the ambient cluster gas that was pushed aside during its passage will fall back and produce tail shocks.

Since no weak lensing analysis is available as yet, nothing can be said about the possible role of collisionless dark matter.

\begin{figure}[htbp]
   \includegraphics[width=13cm]{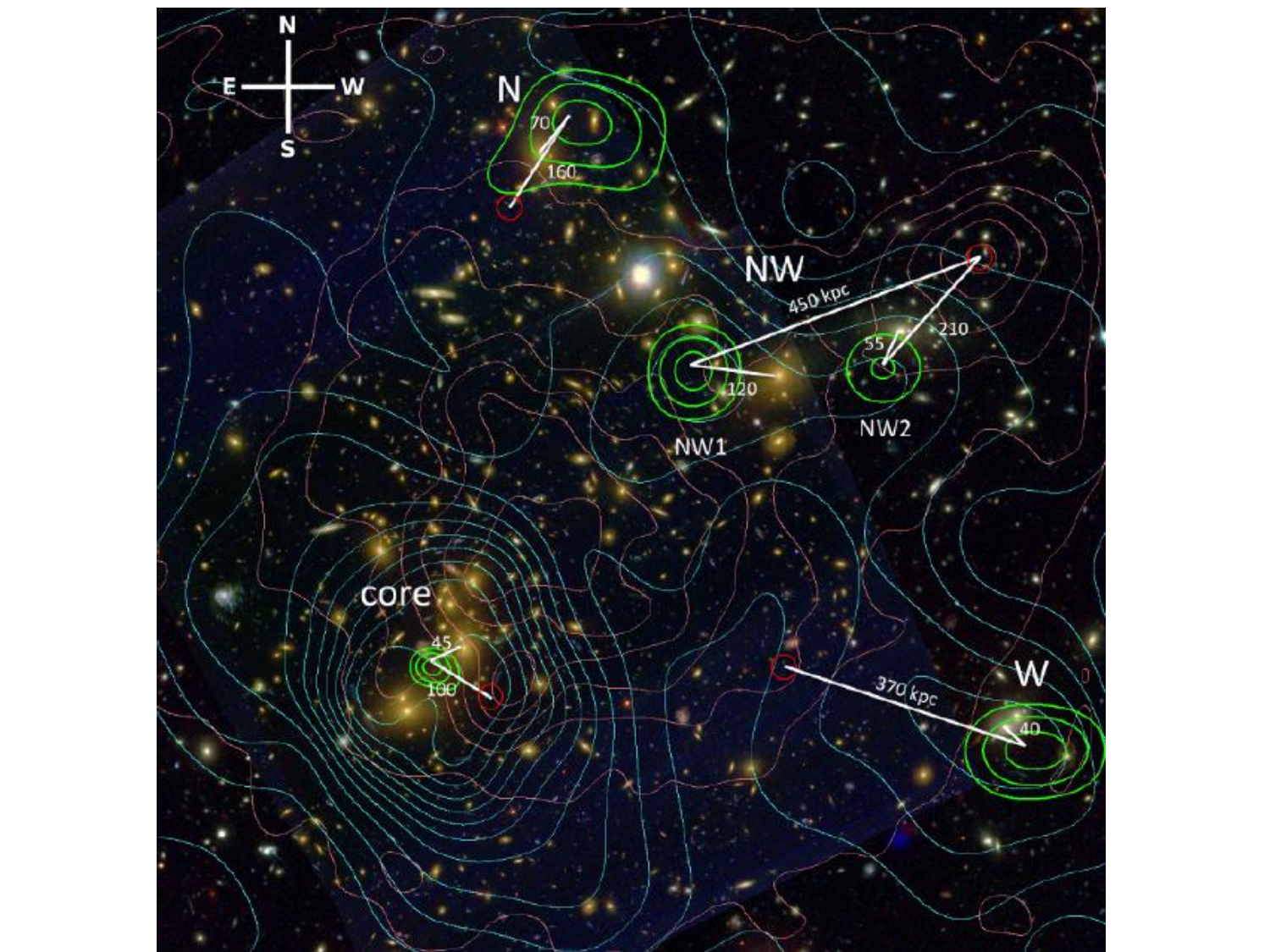}
   \caption{The surface-mass density contours of the merging cluster Abell 2744 are shown in cyan and the X-ray luminosity contours in magenta. The peak positions of the Southern core, the N, NW, and W clumps are indicated by the green likelihood contours. The small red circles show the positions of the local overdensities in the gas distribution, associated with each DM clump. The white rulers show the separation between DM peaks and the bright cluster galaxies and local gas peaks. From J. Merten \& $al$. \cite{Merten}.}
\end{figure}

\subsection{The merging cluster Abell 2744}
Newly acquired data with the \textit{Advanced Camera for Surveys} on the \textit{Hubble Space Telescope, HST,} shows that the cluster Abell 2744 is a complicated merger between three or four separate bodies, as analyzed by J. Merten \& $al$. \cite{Merten}. The position and mass distribution of the Southern core have been tightly constrained by the strong lensing of 11 background galaxies producing 31 multiple images. The N and NW clumps lack such images from strong lensing, indicating that they are less massive. There is also weak lensing information from \textit{HST, VLT}, and \textit{Subaru} available.

The joint gravitational lensing analysis combines all the strongly lensed multiply-imaged systems and their redshifts with weak lensing shear catalogues from all three telescopes to reconstruct the cluster's lensing potential, shown in Fig. 22. The Core, NW and W clumps are clear detections in the surface-mass density distribution with $11\sigma,~4.9\sigma$ and $3.8\sigma$ significance over background, respectively. Somewhat fainter with $2.3\sigma$ significance is the N structure, but it clearly coincides with a prominent X-ray substructure found by M. S. Owers \& $al$. \cite{Owers}.

To determine the geometric configuration of the collision, the location of shock fronts and velocities, densities and temperatures in the intracluster medium, all existing X-ray data from \textit{Chandra} \cite{Owers} were included and reanalyzed. Overlaying the lensing mass reconstruction and the luminosity contours of the emission in Fig. 23 shows an extremely complex picture of separations between the dark matter and baryonic components.

In the core region which is the most massive structure within the merging system, there is no large separation between the distributions of total mass and baryons. The separation of the peaks in the lensing and X-ray maps is similar to that in the \textit{Bullet cluster} \cite{Markev04} and \textit{Baby Bullet} \cite{Bradac}. The Northern mass substructure is $\approx 2.6$ times lighter than the Core, and the X-ray emission lags behind the dark matter to the South.

\begin{figure}[htbp]
   \includegraphics[width=13cm]{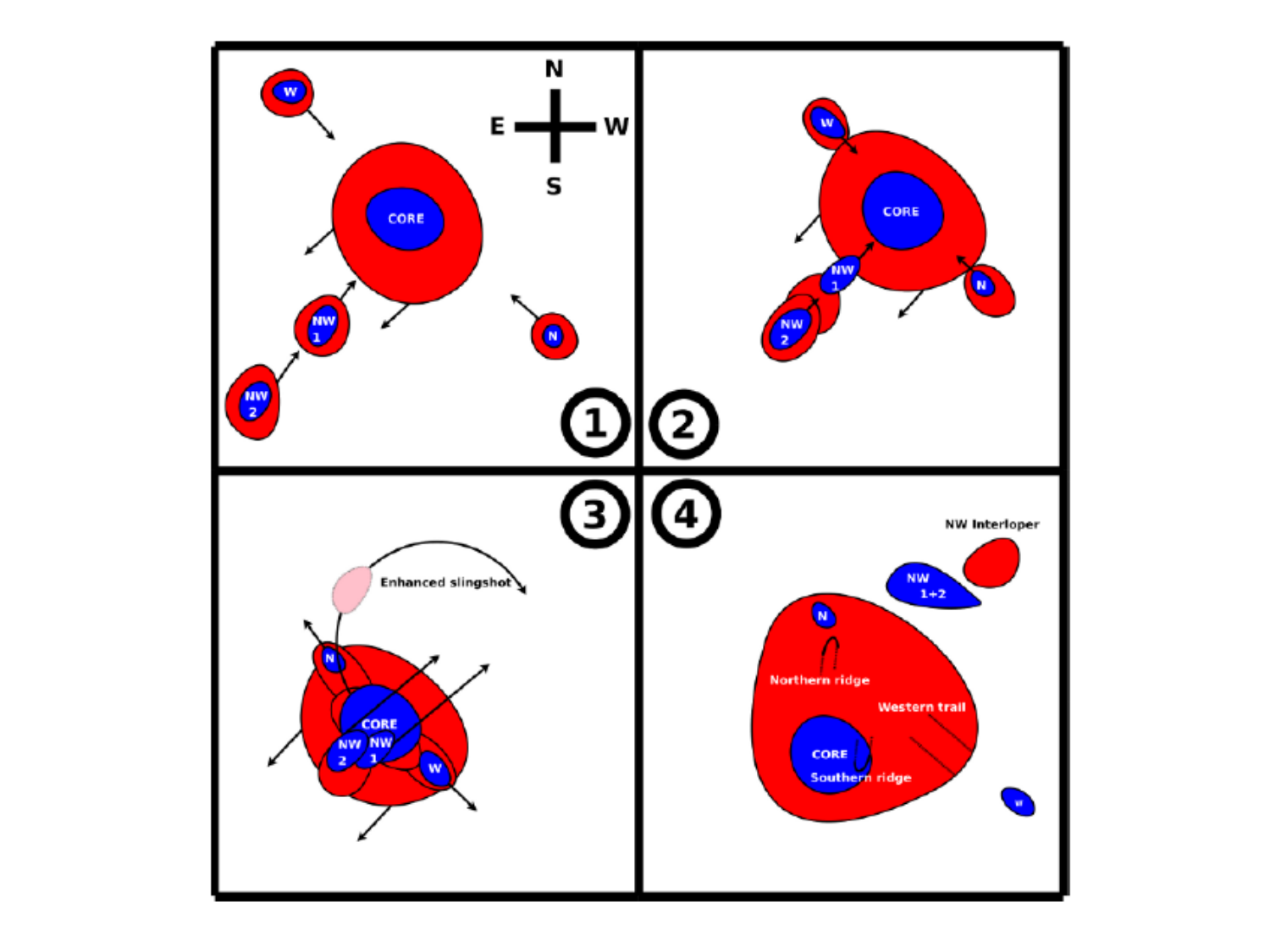}
   \caption{The proposed merger scenario of the cluster Abell 2744 in time-ordered sequence. The NE-SW subclusters merge first (1) with the core, followed very soon (2) by the second merger, in the SE-NW direction. The gas slingshots (3) away to its present position at the extreme NW. In (4) we see the present setup. From J. Merten \& $al$. \cite{Merten}.}
\end{figure}
The substructure in the Northwest is the second most massive and there might also be a second peak in the more Western area of the NW mass clump. However, it is difficult to ascertain whether this is a single, separate DM structure and to derive decisive separation between DM, X-ray luminous gas and bright cluster member galaxies. The X-ray peak to the Northwest of the NW2 mass peak appears to be an X-ray feature with no associated matter or galaxies, a "ghost" clump.

One possible interpretation \cite{Merten} of the complex merging scenario that has taken place in Abell 2744 is a near simultaneous double merger $0.12-0.15$ Gyr ago. first in the NE-SW direction, cf. Fig. 23. The Western clump probably passed closest through the main cluster, as it had its ICM ram-pressure stripped completely. The second merger, in the SE-NW direction, could even have consisted of two small clumps falling into the core, attracted by the core and the Northern and Western clumps. After a first core passage, gas initially trails its associated DM but, while the dark matter slows down,the gas slingshots past it due to a combination of low ram-pressure stripping and adiabatic expansion and cooling \cite{Owers}, ending up as the "ghost" clump. This scenario still requires further observations as well as verification via numerical simulations.

\begin{figure}[htbp]
   \includegraphics[width=13cm]{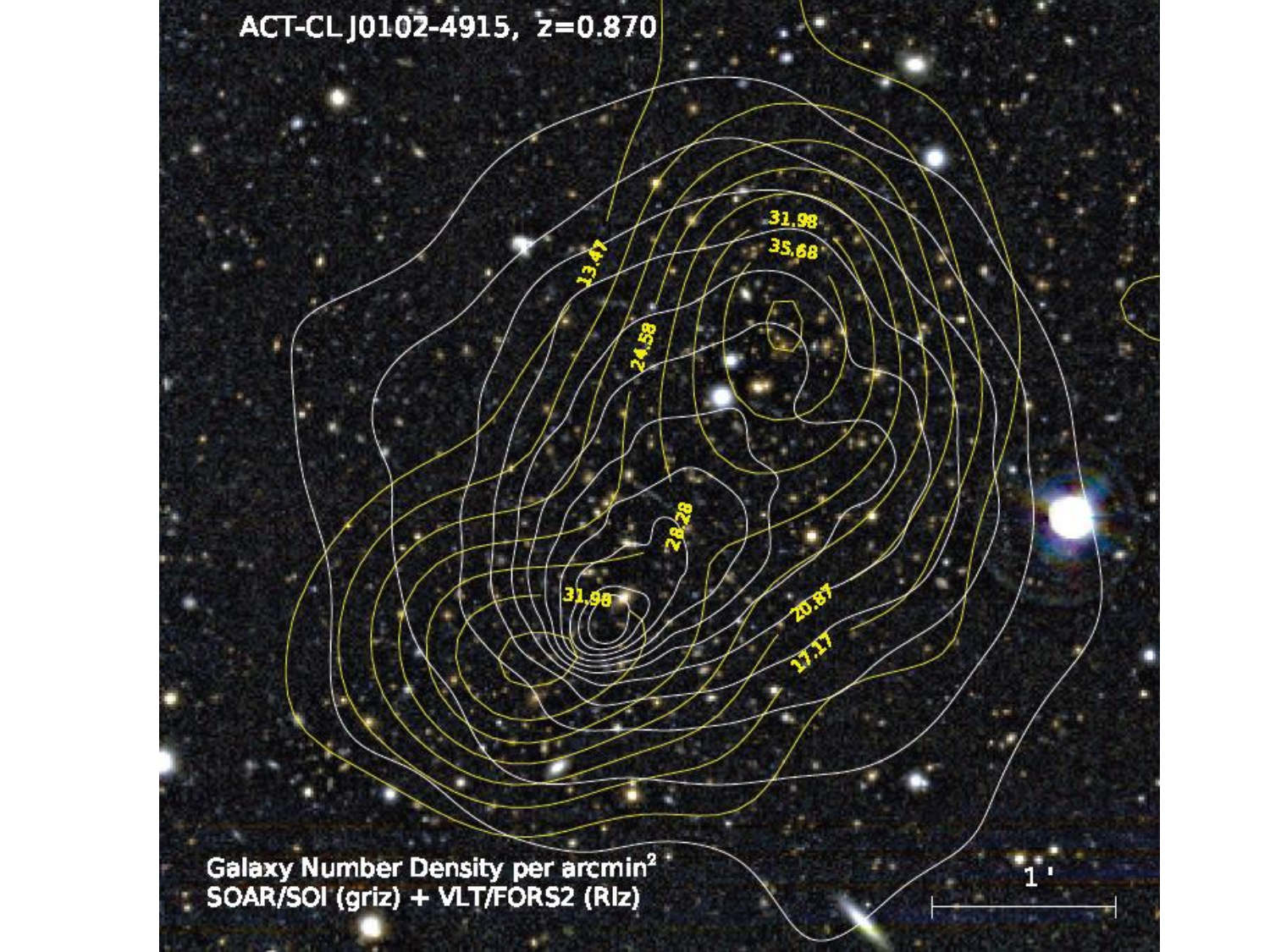}
   \caption{Isopleth contours in yellow of the number density of galaxies in the cluster ATC-CL J0102-4915. The white contours show the X-ray emission. From F. Menenteau \& $al$. \cite{ElGordo}}
\end{figure}

\subsection{"El Gordo", the fat cluster ACT-CL J0102-4915}

The Atacama Cosmology Telescope has presented properties for an exceptionally massive merging cluster, the ACT-CL J0102-4915 nicknamed \textit{El Gordo} at redshift $z=0.87$. It was
discovered by Marriage \& al. \cite{Marriage} selected by its bright Sunyaev-Zel'dovich (SZ) effect, confirmed optically and through its \textit{Chandra} X-ray data \cite{ElGordo}. It is the most significant SZ cluster detection to date by nearly a factor of two, with an  SZ decrement comparable to the \textit{Bullet} cluster 1E0657-558 \cite{Bradac06}.

As can be seen from Fig. 24, the galaxy distribution is double peaked, whereas the peak in the X-ray emission lies between the density peaks. The X-ray peak forms a relatively cool bullet of low entropy gas like in the 1E0657-558. The steep fall-off in the X-ray surface brightness towards the SE, as well as the ``wake" in the main cluster gas toward the NW, indicate that the bullet is apparently moving toward the SE.
The SZ and X-ray peaks are offset similar to that reported for the bullet-like cluster Abell 2146 \cite{Carmen}.

In the absence of a weak lensing mass reconstruction, the galaxy distribution can only be used as a proxy for the total mass distribution. Thus to conclude that an offset between baryonic and DM has been demonstrated is yet premature.

\subsection{The cluster merger DLSCL J0916.2+2951}
A newly discovered \cite{Dawson} major cluster merger at z = 0.53 is DLSCL J0916.2+2951, in which the collisional cluster gas has become clearly dissociated from the collisionless galaxies and dark matter. The cluster was identified using optical and weak-lensing observations as part of the Deep Lens Survey. Follow-up observations with Keck, Subaru, Hubble Space Telescope, and Chandra show that the cluster is a dissociative merger which constrain the DM self-interaction cross-section to $\sigma m_{DM}^{-1}\leq 7$ cm$^2$/g. The system is observed at least $0.7\pm 0.2$ Gyr since first pass-through, thus providing a picture of cluster mergers 2-5 times further progressed than similar systems observed to date.

\section{Comments and conclusions}
What we have termed ``dark matter'' is generic for observed gravitational effects
on all scales: galaxies, small and large galaxy groups, clusters and superclusters, CMB anisotropies over the full horizon, baryonic oscillations over large scales, and cosmic shear in the large-scale matter distribution. The correct explanation or nature of dark matter is not known, whether it implies unconventional particles or modifications to gravitational theory. but gravitational effects have convincingly proved its existence in some form.

The few per cent of the mass of the Universe found as baryonic matter in stars and dust clouds is well accounted for by nucleosynthesis. If there exist particles which were very slow at time $t_{\rm eq}$ when galaxy formation started, they could be candidates for cold dark matter. They must have become non-relativistic much earlier than the leptons, and then decoupled from the hot plasma.

Whenever laboratory searches discover a new particle, it must pass several tests in order to be considered a viable DM candidate: it must be neutral, compatible with constraints on self-interactions (essentially collisionless), consistent with Big Bang nucleosynthesis, and match the appropriate relic density. It must be consistent with direct DM searches and gamma-ray constraints, it must leave stellar evolution unchanged, and be compatible with other astrophysical bounds.

The total dynamical mass of an astronomical system is derivable from the velocity dispersions or the rotation velocities of its components via the use of the Virial Theorem or Kepler's law, respectively. A most important probe is strong gravitational lensing which measures the total mass, but also weak lensing, the oscillations in the Cosmic Microwave Background and in the ambient baryonic medium. Probes separating dark matter from total matter require in addition observations of visible light, infrared radiation, X-rays, the Sunyaev-Zel'dovich effect, and supernovae. Depending on the system under study there are many ways to combine these tools using empirical halo models, simulating stellar population models and galaxy formation models, comparing mass-to-light ratios and mass autocorrelation functions. The most remarkable systems are merging galaxy clusters which, by their motion, separate non-collisional dark matter from optically visible galaxies and hot, radiating gas.

Regardless of the nature of dark matter, all theories attempting to explain it share the burden to explain the gravitational effects described in here. Thus there remains much to be done.

\section*{Acknowledgements}
I am grateful to Sylvain Fouquet, Carmen Rodr\'{\i}guez-Gonz\'{a}lvez, and Will Dawson for clarifying comments and addenda.

\end{document}